\documentclass[reprint,superscriptaddress,amsmath,amssymb,aps,prl,floatfix]{revtex4-1}

\usepackage{amsmath,amssymb,graphicx,mathrsfs,amsfonts}
\usepackage{floatrow}

\newcommand{\ket}[1]{\ensuremath{\left|  #1 \right\rangle}}

\draft 

\begin{document}

\title{Cavity Mediated Collective Spin Exchange Interactions in a Strontium Superradiant Laser} %


\author{Matthew A. Norcia}
\affiliation{JILA, NIST, and Dept. of Physics, University of Colorado, 440 UCB, Boulder, CO  80309, USA}
\author{Robert J. Lewis-Swan}
\affiliation{JILA, NIST, and Dept. of Physics, University of Colorado, 440 UCB, Boulder, CO  80309, USA}
\affiliation{Center for Theory of Quantum Matter, University of Colorado, Boulder, CO 80309, USA}
\author{Julia R.K. Cline}
\affiliation{JILA, NIST, and Dept. of Physics, University of Colorado, 440 UCB, Boulder, CO  80309, USA}
\author{Bihui Zhu}
\affiliation{JILA, NIST, and Dept. of Physics, University of Colorado, 440 UCB, Boulder, CO  80309, USA}
\affiliation{ITAMP, Harvard-Smithsonian Center for Astrophysics, Cambridge, MA 02138, USA}
\author{Ana M. Rey}
\affiliation{JILA, NIST, and Dept. of Physics, University of Colorado, 440 UCB, Boulder, CO  80309, USA}
\affiliation{Center for Theory of Quantum Matter, University of Colorado, Boulder, CO 80309, USA}
\author{James K. Thompson}
\affiliation{JILA, NIST, and Dept. of Physics, University of Colorado, 440 UCB, Boulder, CO  80309, USA}

\date{\today}

\begin{abstract}
{\bf 
\noindent 
Laser cooled and quantum degenerate atoms are widely being pursued as quantum simulators that may explain the behavior of strongly correlated material systems, and as the basis of today's most precise sensors. 
A key challenge towards these goals is to understand and control coherent interactions between the atoms. 
Here, we observe long-range exchange interactions mediated by an optical cavity, which manifest as tunable spin-spin interactions on the pseudo spin-1/2 system composed of the millihertz linewidth clock transition in strontium.  We observe the so-called one axis twisting dynamics, the emergence of a many-body energy gap, and signatures of gap protection of the optical coherence against certain sources of decoherence. These effects manifest  in the output of a pulsed, superradiant laser operating on the millihertz linewidth transition.  Our observations
will aid in the future design of versatile  quantum simulators that take advantage of the unique control and probing capabilities of cavity QED  and the rich internal structure of long-lived Sr atoms. They also open  a route for the next generation of atomic clocks that utilize quantum correlations for enhanced  metrology.}
\end{abstract}


\maketitle
\begin{figure*}[!htb]
\includegraphics[width=6.5in, ]{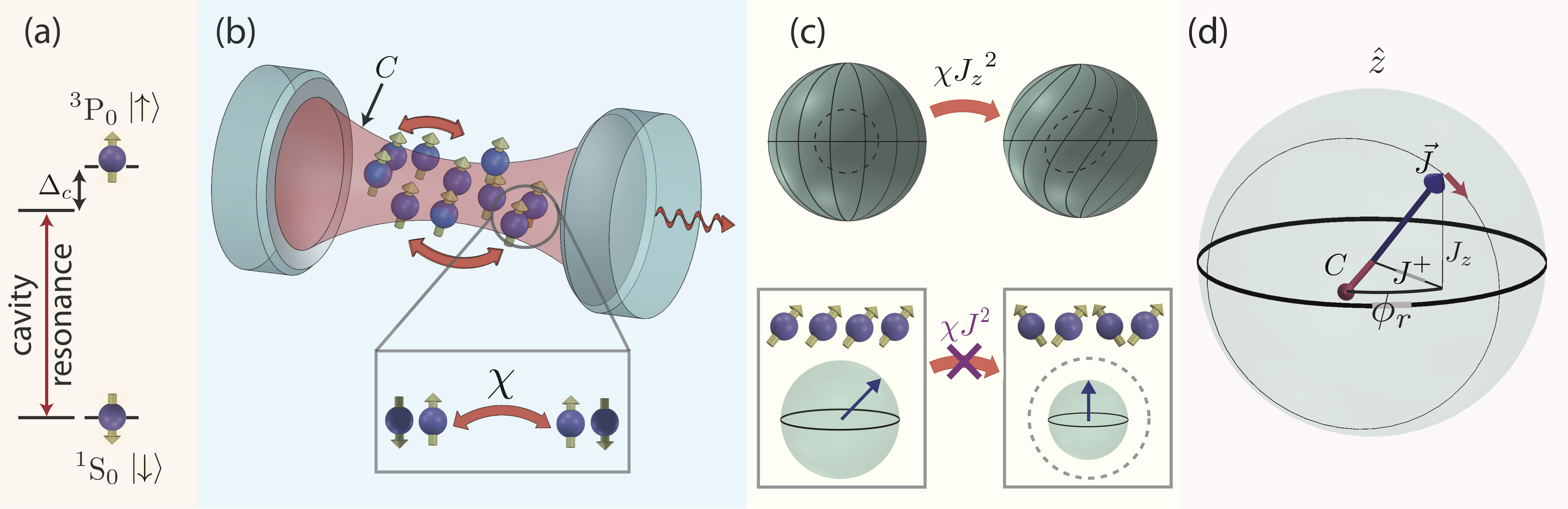}
\caption{(a) An ensemble of $^{87}$Sr atoms interacts with a detuned mode of a high-finesse optical cavity that couples to the millihertz linewidth (150~s lifetime) $^1$S$_0$ to $^3$P$_0$ optical clock transition.  
(b) The cavity mode mediates interactions between the atoms, which lead to both dissipative dynamics in the form of superradiant emission of light through the cavity mirrors and to unitary spin-exchange dynamics that are governed by a Hamiltonian of the form $\hat{H}_{\mathrm{eff}} = \hbar \chi[\hat{J}^+\hat{J}^-]$.  Exchange interactions cause one atom to emit a photon which is then absorbed by another atom, driving anti-correlated spin flips.
(c) The spin-exchange interaction can be rewritten as $\hat{H}_{\mathrm{eff}} \approx \hbar \chi[\hat{J}^2 - \hat{J}_z^2]$. The $\chi \hat{J}_z^2$ term in $\hat{H}_{\mathrm{eff}}$ leads to an inversion dependent frequency shift known as one-axis twisting (OAT), while the $\chi \hat{J}^2$ creates a many-body energy gap  that suppresses detrimental changes in the total spin caused by single particle dephasing.  
(d)  The cavity-mediated dynamics in our experiment can be described at the mean-field level.  The collective Bloch vector $\vec{J}$ rotates about the cavity field vector $C$, which rapidly adjusts to follow the atomic coherence $J^+$ up to fixed angle $\phi_r = \mathrm{tan}^{-1} (2\Delta_c/\kappa)+\pi/2$.  When the cavity is near resonance, this leads primarily to a rotation of the Bloch vector from the north pole toward the south pole (superradiance), while in a far off-resonance cavity, the rotation  primarily causes a horizontal displacement of the Bloch vector which manifests as a shift in the apparent atomic transition frequency.  Because the phase of the cavity field is locked to the phase of the atomic coherence, light exiting the cavity provides us with a real-time probe of the atomic dynamics.  }
\label{fig:diagrams}
\end{figure*}


A crucial requirement for the development of atomic quantum simulators is the ability to create controllable coherent interactions between the atoms.  Implementations of these interactions include direct atomic collisions \cite{Bloch2008r}, direct electric and magnetic dipole interactions \cite{Saffman2010,Labuhn2016,Gross2017, Lahaye2009,dePaz2013, Moses2017}, phonon-mediated couplings in trapped ions \cite{Kim2010,Britton2012,Barreiro2011}, and photon-mediated coupling in a driven optical cavity \cite{Leroux10, BGB2010}.  In this work, we introduce a new type of interaction to this list: spin-exchange interactions between ultra-long-lived optical dipoles mediated by photons in an undriven optical cavity. The effective spins are encoded in the  ground and excited state of the millihertz linewidth strontium clock transition (see Fig.~\ref{fig:diagrams}a,b). This optical transition currently forms the basis of the most precise atomic clocks \cite{Nicholson2015,Bloom2014}, and is a promising candidate for the development of superradiant optical lasers with coherence times beyond 100 seconds \cite{MYC09, Norciae1601231}.  

The exchange interactions manifest in our system as a  collective XX-Heisenberg spin model,  an iconic model  that describes the behavior of a broad class of phenomena ranging from superconductivity \cite{Anderson2008} to quantum magnetism \cite{Auerbach1994}.  We observe evidence of two of the main characteristic features of the collective XX- Heisenberg  model dynamics (see Fig.~\ref{fig:diagrams}c): an orientation-dependent global spin precession of the collective Bloch vector, referred to as one-axis twisting (OAT) \cite{Kitagawa1993}, and the emergence of a many-body energy gap between states of different symmetry \cite{martin2013quantum}. One-axis twisting can generate useful spin squeezing \cite{Kitagawa1993}, Schrodinger cat states \cite{Molmer1999}, quantum phase magnification \cite{Hosten1552}, and enables new measures of entanglement \cite{Garttner2017}.  The energy gap can protect collective dynamics against single-particle sources of dephasing that limit the atomic coherence times needed for high precision measurements and for preserving interesting quantum correlations \cite{Rey2008a}.  

OAT dynamics have been used to generate metrologically useful entanglement in driven systems \cite{Leroux10, bohnet2016quantum},  have been observed in microwave atomic clocks \cite{bize2001cavity} and have independently been proposed as a method to generate entanglement in an undriven cavity system similar to ours \cite{Hu2017,Borregaard2017}.  Spin-locking effects generated by exchange interactions \cite{Laloe1982, Bashkin1986} have been observed in  a variety of  physical systems including  NMR experiments in spin polarized hydrogen \cite{Johnson},$^3$He-$^4$He mixtures \cite{Gully}, and  in trapped cold atoms \cite{McGuirk2002,Deutsch2010, Du2008,Kleine2011}.  In all of these systems, the exchange interactions emerge from  direct collisions.  In contrast, these effects emerge in our system from the exchange of optical photons in a cavity, which makes them decoupled from atomic motion and fully tunable by controlling the cavity parameters.

The cavity-mediated interactions lead both to the collective enhancement of photon emission, as previously demonstrated in our system \cite{Norciae1601231}, and to unitary spin dynamics that emerge when the optical cavity is tuned off resonance from the radiating transition.  This system provides both a new platform for quantum simulation of multi-level spin models, and a potentially powerful tool for precision frequency metrology.  For both of these applications, it is crucial to understand the nature of the cavity-mediated interactions in the system.


Our  experimental system, described in detail in \cite{Norciae1601231}, consists of roughly $N=10^5$ $^{87}$Sr atoms cooled to 10~$\mu$K and tightly trapped by a deep one dimensional optical lattice that is supported by an optical cavity and generates the same confinement for the ground $\ket{\downarrow} \equiv ^1$S$_0$ and excited $\ket{\uparrow} \equiv ^3$P$_0$ clock states. The atoms couple to the cavity via the clock transition with a single-photon Rabi frequency of up to $2g \simeq 2 \times 2\pi \times 4$~Hz.  Near the clock transition wavelength $\lambda_c=698$~nm, the cavity has a finesse of $F = 2.4 \times 10^4$ and a linewidth of $\kappa = 2\pi \times 160$~kHz.  In practice, $g$ varies between  lattice sites, but this can be accounted for by renormalization of parameters (See supplementary material). The cavity detuning from the atomic transition frequency, $\Delta_c$, is varied to generate  exchange interactions. 


We operate in the bad cavity limit where the cavity photons decay much faster than the atomic coherence. 
In this regime photons leaking out of the cavity lead to dissipation in the form of collective atomic decay (superradiance), which  can be described by the collective jump operator $\sqrt{\Gamma/2} \hat{J}^-$,  with $\Gamma = 4g^2\kappa/(4\Delta_c^2 + \kappa^2)$. The collective spin operators $\hat{J}^\pm=\hat{J}_{x}\pm i \hat{J}_{y} $,   characterize the atomic coherence between the ground and excited states. Here $\hat{J}_{x,y,z}=\frac{1}{2}\sum_{i=1}^N \hat \sigma_{i}^{x,y,z} $ and $\hat \sigma_{i}^{x,y,z}$ are Pauli operators acting on the $\ket{\uparrow}$ and $\ket{\downarrow}$ clock state of atom $i$.

\begin{figure*}[!htb]
\includegraphics[width=5in, ]{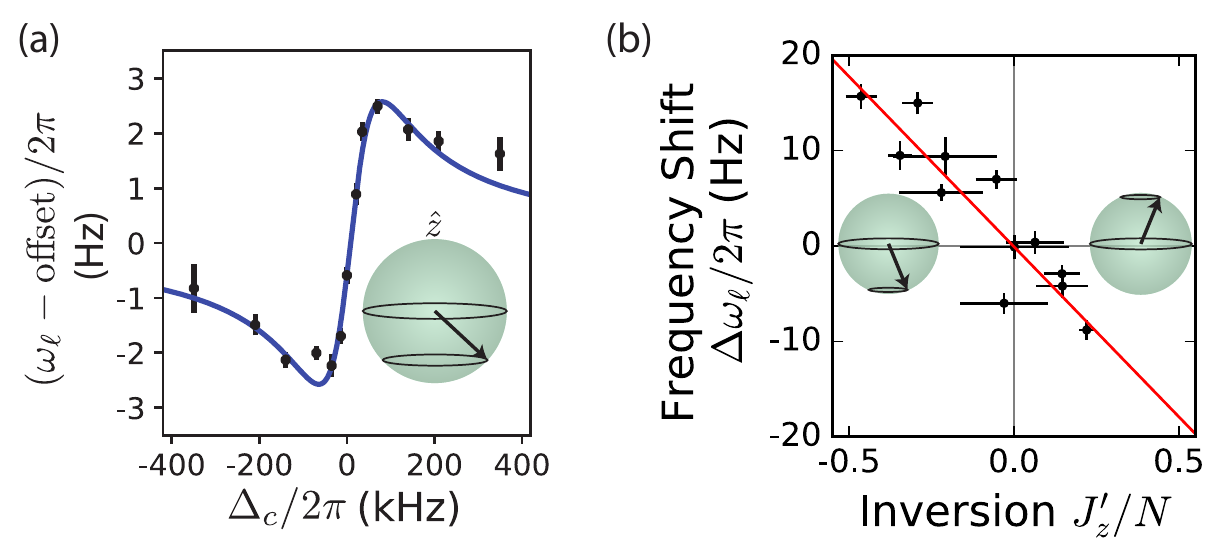}
\caption{Experimental observation of OAT dynamics.  (a) The measured frequency of the emitted light $\omega_\ell$ (with an arbitrary offset subtracted) versus cavity detuning $\Delta_c$ shows the expected dispersive behavior for atoms prepared below the equator of the Bloch sphere.  The blue line is a fit with cavity linewidth held to its independently measured value.  (b)  The measured frequency shift of the emitted laser light $\Delta \omega_\ell \equiv \omega_\ell|_{+29~ \mathrm{kHz}}-\omega_\ell|_{-29~\mathrm{kHz}}$ when the cavity is tuned between $\Delta_c/2 \pi = \pm 29$~kHz shows a linear dependence on effective atomic inversion $J_z^\prime$ .  A simple linear fit to the data (red) yields a ratio of the measured shift to the predicted value based on known cavity parameters, atomic properties and atom number of 0.7(3). We provide a more careful theoretical modeling of the experiment in the supplemental material.  
}
\label{fig:shifts}
\end{figure*}

As we detune the cavity from resonance, interesting cavity-mediated unitary dynamics emerge.  At the microscopic level, the dynamics correspond to a spin-exchange process, where one atom may emit a photon into the cavity that is then absorbed by another.  However, because we have control over and access to collective operators, it is useful to describe the dynamics using an effective XX-Heisenberg Hamiltonian written in terms of collective operators as

\begin{equation}
\hat{H}_{\mathrm{eff}} = \hbar \chi\hat{J}^+\hat{J}^- \equiv \hbar \chi[\hat{J}^2 - \hat{J}_z^2 + \hat{J}_z]
\end{equation} 

\noindent Here $\chi = 4g^2\Delta_c/(4\Delta_c^2 + \kappa^2)$,  $\hat{J_z}$ corresponds to atomic inversion, and $\hat{J}^2= \hat{J}_x^2+\hat{J}_y^2+\hat{J}_z^2$ is the total spin operator. We have written the Hamiltonian in the rotating frame of the atomic transition frequency $\omega_a$.  Here, we will focus on the second form, which is useful for describing the observed collective dynamics of the system.  The last term $\propto \hat{J}_z$ induces  a small single-particle rotation  and can be safely ignored in the large $N$ limit where our experiment operates.

The two remaining terms are equally important, and generate distinct effects.  The $\chi \hat{J}_z^2$ term realizes one-axis twisting (OAT), illustrated in Fig.~\ref{fig:diagrams}c. The $\chi\hat{J}^2$ term induces a many-body energy gap as large as $2 \chi J$ between adjacent states with total spin $J$ and $J-1$.

The behavior of the Bloch vector components $J_{x,y,z}= \langle \hat{J}_{x,y,z} \rangle$ can be described geometrically at the mean-field (MF) level by treating the cavity optical field as a self-generated effective magnetic field lying in the x-y plane of the Bloch sphere.  This effective field induces a rotation of the Bloch vector about the field's axis at a frequency proportional to the field's magnitude (see  Fig.~\ref{fig:diagrams}d).  
The field can be written in complex notation as $C= C_x + i C_y= (\chi +i \Gamma)J^+$.  
The effective field's azimuthal angle follows the azimuthal angle of the Bloch vector's projection onto the x-y plane $\phi$ up to an offset of $\phi_r$ that is set by the cavity detuning: $\phi_r=\arctan(2\Delta_c/\kappa)+\pi/2$.  By detecting the field leaking out of the cavity, we obtain a real-time non-destructive measure of the phase, frequency, and magnitude $J_\perp=\sqrt{J^+J^-}$ of the transverse component of the Bloch vector $J^+$ \cite{PhysRevA.88.013826}.

At resonance ($\Delta_c=0$), the relative phase is $\phi_r=\pi/2$.  The cavity field $C$ causes a rotation of the Bloch vector from the north pole (all atoms in $\ket{\uparrow}$) to the south pole (all atoms in $\ket{\downarrow}$).  This is the manifestation of collective or superradiant decay in this picture.

When the cavity is tuned away from resonance $\Delta_c\neq 0$, the cavity responds to the drive with either a phase advance or lag.  In the large detuning limit, $|\Delta_c|\gg \kappa/2$, the relative azimuthal angle of the effective field is now $\phi_r =0$ or $\pi$, depending on the sign of the detuning.  In this limit, the cavity field generated by the atoms drives a rotation of the Bloch vector that changes its azimuthal angle, but not its polar angle.  We interpret this additional precession of the Bloch vector's azimuthal angle as an inversion-dependent frequency shift $\omega_{OAT}=-2 \chi J_z$ of the atomic coherence: ${J}^+\sim{J}^+(0) e^{i\omega_{OAT} t}$.  This same frequency shift is inherited by the light emitted from the cavity.  


To generate a spin-coherent atomic  state, we optically pump the atoms to $^1$S$_0$, $m_F = 9/2$, and coherently drive the  $^1$S$_0$ to $^3$P$_0$ transition through the optical cavity with light polarized to maintain spin projection $m_F$.  This rotates the atomic Bloch vector up from the south pole of the Bloch sphere.  By changing the duration or amplitude of this coherent drive, we may prepare a state with arbitrary inversion.  The exact details of this state preparation depend on the standing-wave nature of the cavity mode, as is described in the supplementary materials.  

After switching off the coherent drive, we overlap the subsequently emitted superradiant light with a very stable reference laser to form a heterodyne beat note to determine the light's phase and it's frequency $\omega_\ell$.  We extract $\omega_\ell$ from only the first 8~ms of the superradiant pulse, during which, changes in inversion are small.    

First, we explore the variation of $\omega_\ell$ with cavity detuning $\Delta_c$ (Fig.~\ref{fig:shifts}a) at fixed initial inversion. On each trial, we prepare the atoms in the same state below the equator of the Bloch sphere.  We observe the expected dispersive behavior $\omega_\ell \propto \Delta_c/(4\Delta_c^2 + \kappa^2)$ of the frequency shift with detuning, as can be seen by the fitted dispersive curve  with the cavity linewidth held fixed to its independently measured value.

Fig.~\ref{fig:shifts}b displays a roughly linear scaling of the frequency shift versus an effective population inversion $J_z^\prime$, that accounts for inhomogeneous coupling of atoms to the cavity (see supplemental materials). 
To cancel technical noise, we plot the measured change in frequency of the emitted light $\Delta\omega_\ell /2 \pi$ when the cavity is detuned by $\Delta_c/2 \pi = \pm 29$~kHz. 
This inversion-dependent frequency shift $\propto J_z^\prime$ is the manifestation of OAT dynamics at the mean-field level.  In the supplemental material we provide a more detailed theoretical modeling of the experiment that explains some of the spread in the data points.  Specifically, the model accounts for the impact of the variation of $J_z^\prime$ during the measurement window, a variation that depends on the exact details of the initial state preparation.


So long as $J$ is constant, the distinction between our exchange Hamiltonian and a simpler OAT Hamiltonian is irrelevant for our observables.  However, when non-collective effects are present that would modify $J$, the $\chi \hat{J}^2$ term in the Hamiltonian significantly modifies the dynamics.  

\begin{figure}[!htb]
\includegraphics[width=3.375in, ]{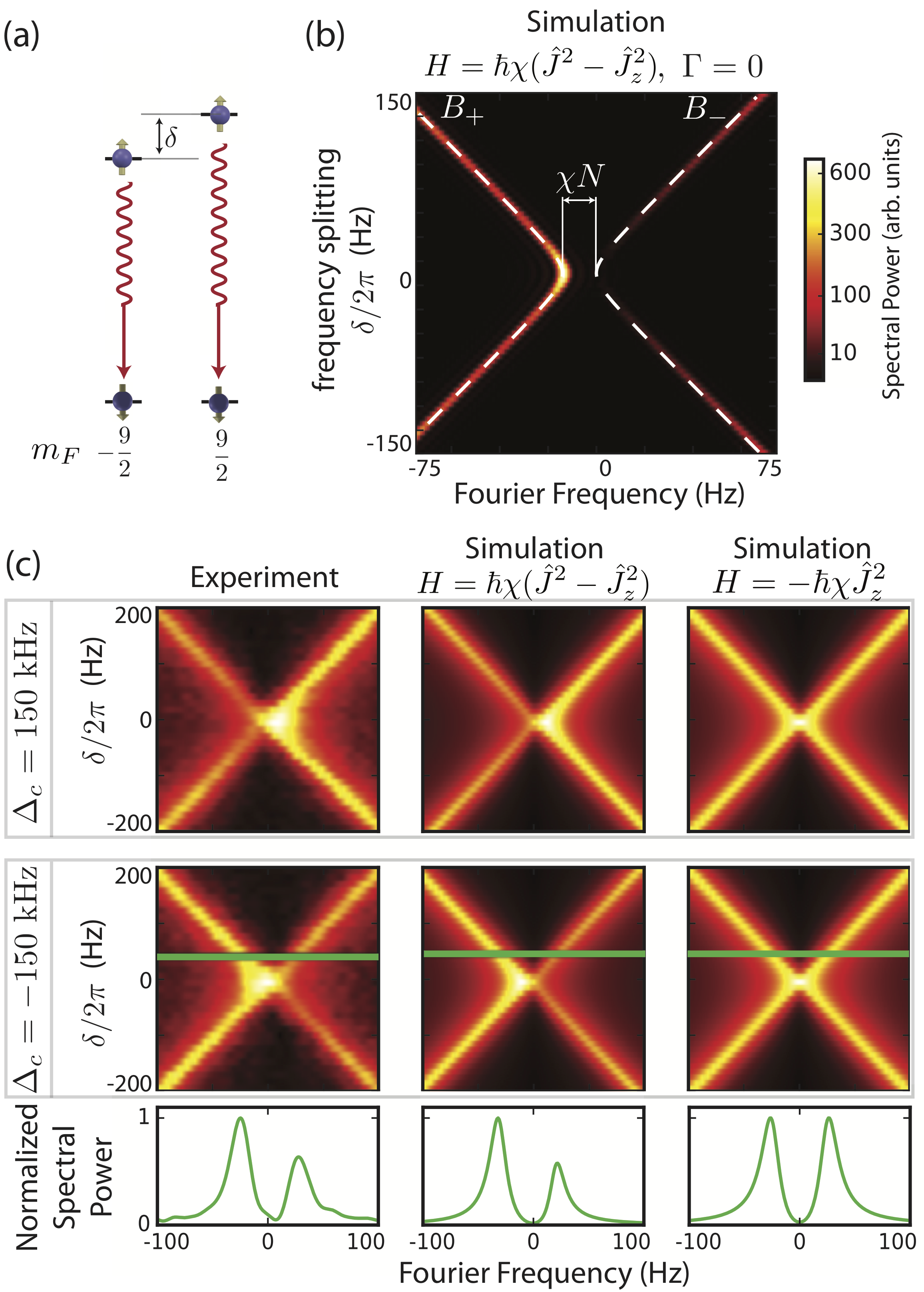}
\caption{(a) We introduce controlled inhomogeneity by simultaneously populating the $m_F = \pm 9/2$ states and applying a magnetic field to split the Zeeman sublevels by a frequency $\delta$.  
(b) A simulation of radiated spectra versus $\delta$ for states prepared near the south pole of the Bloch sphere, in the absence of dissipation ($\Gamma = 0$).  The gap protection term in the Hamiltonian, ($\hbar \chi\hat{J}^2$) leads to a frequency splitting $\chi N$ of normal modes $B_\pm$ (dashed lines) at small $\delta$, and an imbalance in power radiated from the two modes.  
(c) Comparison of experimental results to theory that includes dissipation for atoms prepared near the south pole of the Bloch sphere.  We observe hints of a normal mode splitting and a clear asymmetry in radiated power that switches direction between cavity detunings of $\Delta_c/2\pi = 150 \ \mathrm{kHz} \ \simeq \kappa/2 \pi$ (top row) and $\Delta_c/2\pi = -150 \ \mathrm{kHz}$ (middle row) in both experiment (left column) and a simulation of $\hat{H} = \hbar \chi[\hat{J}^2 - \hat{J}_z^2]$ (center column).  A simulation of  the pure OAT Hamiltonian $\hat{H} = \hbar \chi\hat{J}_z^2$ (right column) does not reproduce these features, highlighting the role of the gap protection term in the observed spectra.  Power spectra taken at a fixed value of $\delta$ indicated by horizontal green line (bottom row) highlight this comparison.
}
\label{fig:jdotj}
\end{figure}

To observe these modifications, we prepare half of the atoms in $m_F=-9/2$ (described by a Bloch vector $\vec{J}_1$) and the other half in $m_F=9/2$ (described by a Bloch vector $\vec{J}_2$) \cite{PhysRevLett.113.154101,Weiner2017}, as shown in Fig.~\ref{fig:jdotj}a. The two ensembles experience a differential  Zeeman  energy shift  $\hbar \delta$ proportional to an applied magnetic field.  The Hamiltonian can be written as $\hat{H} = \hbar\chi\hat{J}^+\hat{J}^- +  \hbar\delta \hat{j}_z$ where the sum and difference operators are defined as $\hat{\vec{J}} = \hat{\vec{J}}_1 + \hat{\vec{J}}_2$ and  $\hat{\vec{j}} = \hat{\vec{J}}_1 - \hat{\vec{J}}_2$.


The mean-field equations of motion for the expectation values $J^+$ and $j^+$  (neglecting $\Gamma$  for simplicity) are given by:

 \begin{eqnarray}
  \frac{d J^+}{dt} &=& {-i} {2\chi} J^+(t) J_z(0) {+i}\delta j^+(t) \\\label{eqn:MF2}
   \frac{d j^+}{dt} &=& {-i} {2\chi } J^+(t) j_z(t) {+i}\delta J^+(t)
\end{eqnarray}

\begin{figure*}[!htb]
\includegraphics[width=6.5in, ]{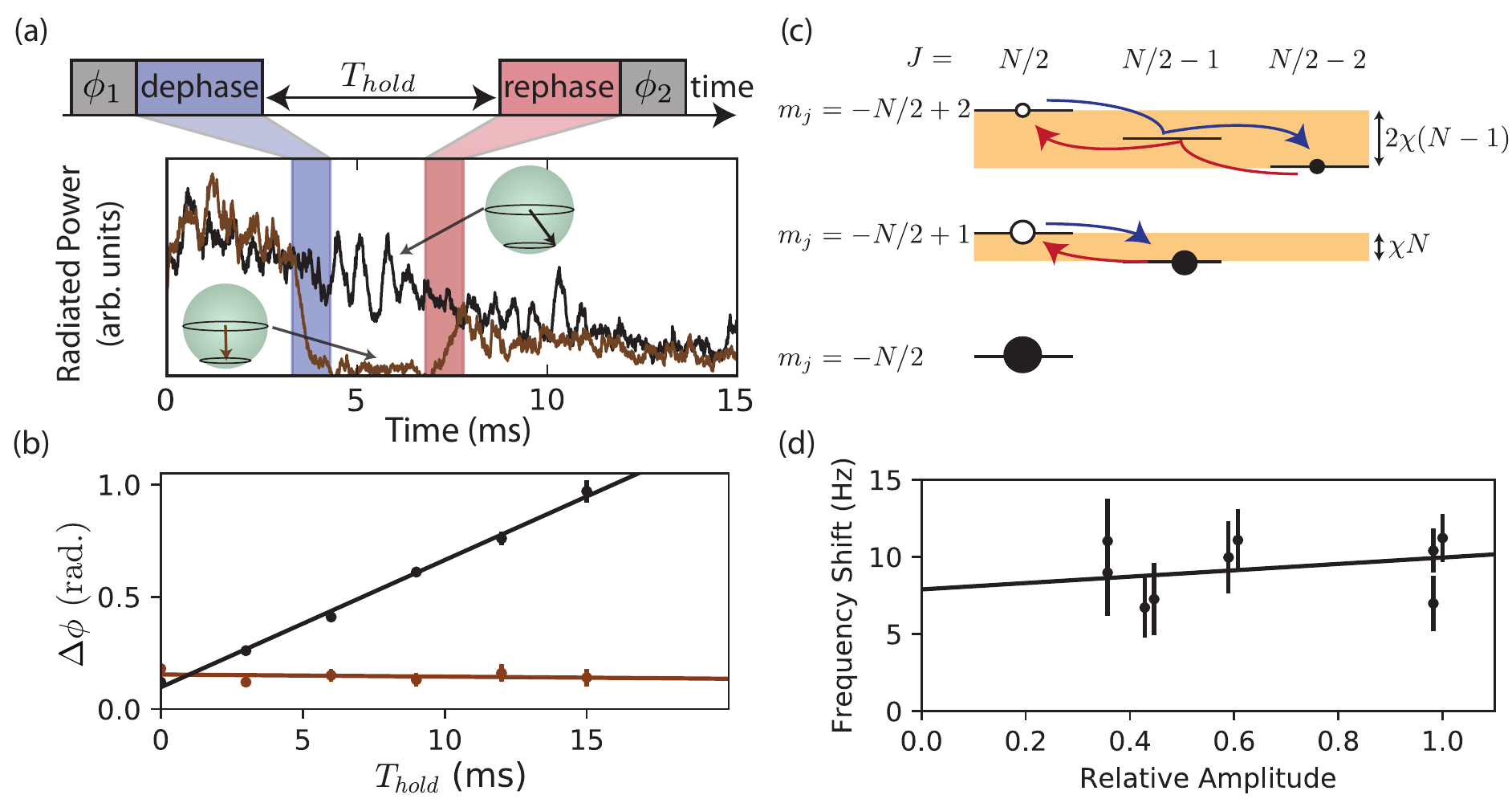}
\caption{Direct spectroscopic observation of the many-body energy gap.  \textbf{(a)} We introduce controlled reversible dephasing to suppress and restore atomic coherence and superradiance before and after a variable time $T_{hold}$.  \textbf{(b)} We measure the change in the difference between optical phase measurements before dephasing ($\phi_1$) and after rephasing ($\phi_2$), as the cavity is toggled between opposite detunings $\Delta_c/2\pi = \pm29$~kHz:  $\Delta \phi  = (\phi_2 - \phi_1)|_{29~\mathrm{kHz}}-(\phi_2 - \phi_1)|_{-29~\mathrm{kHz}}$.  Brown (black) points correspond to trials where the atoms were dephased (not dephased) during $T_{hold}$, corresponding to the dark (bright) normal modes.  $\Delta \phi$ shows no dependence on $T_{hold}$ for the dephased case, and a phase shift that scales linearly with $T_{hold}$ when the atoms are not dephased, confirming a frequency shift between the bright and dark portions of the atomic coherence.
\textbf{(c)} The connection between our observed phase shift and an energy gap can be understood explicitly by considering the energy levels for a Hamiltonian $\hat{H}=  \hbar \chi \hat{J}^2$ in a Dicke basis for states $\ket{J, m_J}$ near $m_J = -N/2$.  States of differing $m_J$ are offset vertically for clarity.  In the experiment, atoms begin in a superposition of maximally symmetric states with $J = N/2$ (far left states).  Dephasing (blue arrows) transfers the atomic ensemble to states of minimal coherence for a given $m_J$, which are shifted in energy relative to the maximally symmetric states by the energy gap.  When coherence is restored in the rephasing step (red arrows), this shift is converted into a phase shift of the atomic coherence, which provides a measurement of the many-body gap (see text for full details).  
\textbf{(d)} Measured shift in $\omega_\ell$ when cavity detuning is toggled $\Delta_c/2\pi = \pm29$~kHz, versus residual emission amplitude when atoms are partially dephased.  A linear fit to the frequency shift versus amplitude returns a slope consistent with zero and an offset at zero amplitude inconsistent with zero shift. }
\label{fig:dephase}
\end{figure*}

\noindent The detuning $\delta$ converts the amplitude of $J^+$ into $j^+$ and back as it causes a relative rotation between the two Bloch vectors $\vec{J}_1$ and $\vec{J}_2$.  
In general, $j_z$ varies with time (see supplemental materials), complicating the interpretation of these equations.  However, when the total Bloch vector is prepared near a pole of the Bloch sphere with initial $j_z=0$ (a case which we consider here both for simplicity and for comparison with experimental observations), energy conservation requires that  $|j_z|$ remains small  $|j_z|/ |J_z|\ll 1 $   and to an excellent approximation $j_z$ can  be set to zero in  Eq.~\ref{eqn:MF2}. 
We can then solve for the normal modes of the system in terms of only $J^+$ and $j^+$.  These are $B_+= \cos( \alpha/2) J^+ +\sin(\alpha/2) j^+$ and $B_-=\sin( \alpha/2) J^+ -\cos(\alpha/2) j^+$, with corresponding frequencies of $\omega_{\pm}=\left(\chi N \pm \sqrt{(\chi N)^2 + 4\delta^2 }\right)/2$ (see Fig.~\ref{fig:jdotj}b). The mode mixing angle $\alpha$ is $\tan \alpha= 2\delta/(\chi N)$. 
The frequency splitting between the two modes when $\delta=0$ is equal to $\chi N$.  This energy gap suppresses the inter-conversion of $J^+$ and $j^+$ when $\delta \ll \chi N$, and is a classical manifestation of the energy gap between states of $J=N/2$ and $J=N/2-1$.  This connection will be made explicit in the discussion of Fig.~\ref{fig:dephase}

In the experiment, we detect the field radiated into the cavity, which is proportional to $J^+$ (and independent of $j^+$).
For small $\delta$, we thus expect that the mode $B_+$ will be bright and the mode $B_-$ will be dim, while for large $\delta$ the two modes should be equally bright.  Further, the two modes should undergo an avoided-crossing type behavior.  These features are clearly apparent in the output of a simulation in which $\Gamma = 0$ (Fig.~\ref{fig:jdotj}b). 

The presence of dissipation $\Gamma$ in the experimentally accessible regime makes quantitative comparison difficult. However, the qualitative signatures of the $\Gamma = 0$ case, especially the imbalanced brightness of the two modes, are still clearly present in the data of Fig.~\ref{fig:jdotj}b.  
Importantly, 
a pure $\chi \hat{J}_z^2$ Hamiltonian leads only to an overall frequency shift of both modes, and cannot explain the apparent curvature near $\delta=0$ nor the dimming of one mode relative to the other, as is shown by simulation that includes dissipation in  Fig.~\ref{fig:jdotj}c.



To measure the frequency splitting associated with the many-body gap, we compare the rate at which bright and dark atomic states accumulate phase.  In the two-spin system explored above, these states could correspond to $B_+$ and $B_-$ respectively when $\delta = 0$. For quantitative measurements however, we found it advantageous to no longer use a two-spin system. Instead, we populate only the $m_f = 9/2$ state and prepare the total Bloch vector  near the south pole of the Bloch sphere in a bright state that radiates light. We can convert this bright state into a dark state by applying inhomogeneous phase shifts to the atoms using a pulse of laser light that is tuned off-resonance from the 7.5~kHz linewidth $^1$S$_0$ to $^3$P$_1$ transition and whose intensity varies over the spatial extent of the atoms.  
The pulse greatly reduces the magnitude of the atomic coherence $J_\perp$, and thus superradiance, while leaving $J_z$ unchanged. The system is then allowed to evolve for a time $T_{hold}$, after which it is reconverted into a bright state by applying a second pulse from the same laser with opposite detuning from the $^1$S$_0$ to $^3$P$_1$ transition, causing the atoms to rephase, and superradiance to be restored.

To access the frequency  of the dark state, we perform two phase measurements $\phi_1$ and $\phi_2$ of the light emitted from the cavity before dephasing and after rephasing, respectively.  We measure how the difference between the  two phases $\phi_2-\phi_1$  change when the cavity is alternately detuned by $\pm 29$~kHz, and label this quantity $\Delta\phi$ (see Fig.~\ref{fig:dephase}).




If the system evolves as a bright state during $T_{hold}$ (i.e.~if we do not apply the dephasing and rephasing pulses ), the measured linear slope of $\Delta\phi$ versus $T_{hold}$ implies that the bright state experiences a frequency shift  of $\pm 4.5$~Hz.  When the system evolves as a dark state during $T_{hold}$ (i.e.~if we apply the dephasing and rephasing steps), the slope of $\Delta \phi$ versus $T_{hold}$ is consistent with no frequency shift.  The difference in frequency between the bright and dark state phase evolutions during $T_{hold}$ is the direct manifestation of the many-body gap for a system near the south pole of the Bloch sphere.  

To make this connection explicit, we may consider the effect of dephasing and rephasing in a Dicke basis for states prepared near $J_z = -N/2$ under the effect of a Hamiltonian $\hat{H}=  \hbar \chi \hat{J}^2$.  We neglect the effects of the $\chi \hat{J}_z^2$ term for now, as this commutes with both $\hat{J}^2$ and $\hat{J}_z$ so we can account for its effects at the end.
States in this basis $\ket{J, m_J}$ are eigenstates of $\hat{J}^2$ and $\hat{J}_z$, and are labelled by the magnitude of their spin $J$ and their projection along $\hat{z}$, $-J \leq m_J \leq J$.  A maximally symmetric coherent spin state $\ket{\theta, \phi}$ with $J=N/2$ (and corresponding Bloch vector with polar angle $\theta$ and azimuthal angle $\phi$) is represented by a superposition of Dicke states as 

\begin{eqnarray}
\ket{\theta, \phi} = \sum_{m_J = -N/2}^{N/2} c(m_J, \theta) e^{- i \phi m_J}\ket{N/2, m_J}
\end{eqnarray}
\noindent where the coefficients $c(m_J, \theta)$ are real.  The constant relative phase between adjacent values of $m_J$ sets the azimuthal angle $\phi$ of the coherent spin state.


The relevant Dicke states for the three lowest values of $m_J$ are shown in Fig.~\ref{fig:dephase}c, along with the energy gaps created by a $\hbar \chi \hat{J}^2$ Hamiltonian.  When ideal dephasing is applied,  Dicke states with constant $m_J$ are coupled such that the maximally symmetric Dicke state \ket{N/2, m_J} is mapped to the the state \ket{|m_J|, m_J}, the Dicke state with the smallest possible value of $J= |m_J|$  compatible with the initial spin projection $m_J$ of each Dicke state.

For Dicke states near the bottom of the Bloch sphere, the energy gap between $\ket{|m_J|, m_J}$ and $\ket{N/2, m_J}$ scales linearly with $m_J$ as $E(m_J)\approx -\hbar \chi N (m_J + N/2)$. After an ideal rephasing step, the phases accumulated during $T_{hold}$ are mapped back from $\ket{|m_J|, m_J}$ to the original Dicke states $\ket{N/2, m_J}$. Because of the time spent in the minimally symmetric states, each Dicke state's amplitude will have acquired a phase shift $\propto \chi N T_{hold} m_J $, equivalent to a change of the coherent state's azimuthal phase $\phi$ by $-\chi N T_{hold}$.  Therefore, we can understand that measuring the difference in phase  accumulated by the Bloch vector during $T_{hold}$ with and without dephasing/rephasing is a direct spectroscopic measure of the many-body energy gap.  

The additional OAT dynamics that we have neglected so far contribute an equal phase shift to both the minimally and maximally symmetric configurations.  This does not effect the difference in frequency between bright and dark states (which is the key signature of the many-body gap), but in our case means that the bright state appears to have a frequency shift, while the dark state deos not.  The fact that we measure a frequency shift of the dark state consistent with zero confirms that the coefficients in front of the $\hat{J}^2$ and $\hat{J}_z^2$ terms in our Hamiltonian are equal and opposite to within our experimental precision, and consistent with a Hamiltonian of the form $\hat{H} = \chi\hat{J}^+\hat{J}^-$.

Instead of the prior procedure, we can instead partially dephase the system and measure the frequency of the residual emission.  
We find that the shift in frequency of the emitted light when the cavity detuning is toggled is independent of the magnitude of the residual coherence, within experimental error (Fig.~\ref{fig:dephase}d).  


Despite the fact that dephasing populates a variable combination of the bright and dark modes of the system, we measured a frequency shift consistent with the bright mode.
This is to be expected, as only the bright mode radiates.  Because $J$ remains fixed during the portion of the experiment over which we measure frequency, we only see the effects of a pure $\chi J_z^2$ Hamiltonian, which do not depend on the degree of atomic coherence.

Our work demonstrates the suitability of ensembles of Sr atoms interacting with an optical cavity via long-lived transitions for the simulation of long-range quantum magnetism. We benchmarked the experimental realization of a collective XX-Heisenberg model finding agreement  with the predictions of mean-field theory.

Our  observations pave a way for future explorations of cavity mediated exchange interactions, which could generate  entanglement and metrologically useful spin squeezing 
while taking advantage of the  spin-locking  mechanism provided  by the many-body energy gap.  The latter can be used to stabilize  rich forms of  strongly  correlated steady states of matter and light  when additionally  driving the system    (incoherently or coherently). 

\section{ACKNOWLEDGMENTS}
We thank the members of Jun Ye's lab for the use their reference light, and Adam Kaufman and Shimon Kolkowitz for useful comments on the manuscript.  All authors acknowledge financial support from DARPA QuASAR, ARO, NSF PFC, and NIST. J.R.K.C. acknowledges financial support from NSF GRFP.
This work was supported by NSF PFC grant number PHY 1734006, DARPA Extreme Sensing, and NIST.

\bibliography{ThompsonLab}

\section{SOM Section 1:  Quantum Description }
The dynamics of the coupled  atom-light system can be described by a master equation for the density matrix, $\hat \rho$,
\begin{equation}
 \frac{d\hat{\rho}}{dt} = -\frac{i}{\hbar}\left[\hat{H}_{\mathrm{AL}},\hat{\rho}\right] + \mathcal{L}_c[\hat{\rho}]  \label{eqn:AL_master_eqn}
\end{equation}
Here, the Hamiltonian describing the atom-light coupling is
\begin{equation}
 \hat{H}_{\mathrm{AL}} = \hbar\Delta_c\hat{a}^{\dagger}\hat{a} - \hbar\sum_{n_l} \frac{\delta_{n_l}}{2} \hat{\sigma}^z_{n_l} + \hbar \sum_{n_l} g_{n_l} \left( \hat{a}^{\dagger}\hat{\sigma}^-_{n_l} + \hat{a}\hat{\sigma}^+_{n_l} \right) 
\end{equation}
where $\Delta_c$ characterizes the relative detuning of the cavity field from the atomic transition, $g_{n_l}$ characterizes the atom-light coupling, $\delta_{n_l}$ describes inhomogeneous shifts in the individual atomic transition frequencies and $\hat{\sigma}^{\alpha}_{n_l}$ 
denote the conventional Pauli matrices with $\alpha=x,y,z$ acting on the clock state's electronic degrees of freedom.
To be completely general, we use notation wherein the subscript $n_l$ indexes all relevant spatial ($l$) and internal degrees of freedom (such as different $m_F$ spin projections). Lastly, the photon loss from the cavity with power decay rate $\kappa$ is described by a Lindblad term,
\begin{equation}
 \mathcal{L}_c[\hat{\rho}] = \frac{\kappa}{2}\left( 2\hat{a}\hat{\rho}\hat{a}^{\dagger} - \hat{a}^{\dagger}\hat{a}\hat{\rho} - \hat{\rho}\hat{a}^{\dagger}\hat{a} \right) 
\end{equation}
whilst single-particle homogeneous broadening of the atomic ensemble is characterised by
\begin{equation}
    \mathcal{L}_{el}[\hat{\rho}_s] = \frac{\gamma_{el}}{2} \sum_{n_l} \left( \hat{\sigma}^z_{n_l}\hat{\rho}_s\hat{\sigma}^z_{n_l} - \hat{\rho}_s \right)  \label{eqn:spin_gammaPerp}
\end{equation}
with strength $\gamma_{el}$. This broadening characterizes dephasing from processes such as background collisions and magnetic field noise in the experiment.

As the cavity loss occurs at a much faster rate than the atomic dynamics, $\kappa \gg g$, we may adiabatically eliminate the cavity mode and describe the atomic system by an effective spin-model. The result is an effective slaving of the cavity 
field to the spin operators,
\begin{equation}
 \hat{a}(t) \equiv \frac{2}{2\Delta_c - i\kappa} \sum_{n_l} g_{n_l} \hat{\sigma}^{-}_{n_l} \label{eqn:SOM_cavitymode}
\end{equation}
leading to a master equation for the reduced density matrix $\hat{\rho}_s$ of the spin degree of freedom \cite{Agarwal_DissipativeMasterEqn_1997}:
\begin{equation}
 \frac{d\hat{\rho}_s}{dt} = -\frac{i}{\hbar}\left[\hat{H}_{\mathrm{eff}},\hat{\rho}_s\right] + \mathcal{L}_{\Gamma}[\hat{\rho}_s] + \mathcal{L}_{el}[\hat{\rho}_s]  \label{eqn:spin_master_eqn}
\end{equation}
with
\begin{eqnarray}
 \hat{H}_{\mathrm{eff}} & = & \hbar\sum_{n_l,m_k} \chi_{n_l,m_k} \hat{\sigma}^+_{n_l} \hat{\sigma}^-_{m_k} - \hbar\sum_{n_j} \frac{\delta_{n_l}}{2} \hat{\sigma}^z_{n_l}  \label{eqn:spin_H} \\
 \mathcal{L}_{\Gamma}[\hat{\rho}_s] & = & \sum_{n_l,m_k} \frac{\Gamma_{n_l,m_k}}{2}\left( 2\hat{\sigma}^-_{n_l}\hat{\rho}_s\hat{\sigma}^+_{m_k}  \right) \notag \\
 & & \left. - \hat{\sigma}^+_{n_l}\hat{\sigma}^-_{m_k}\hat{\rho}_s - \hat{\rho}_s\hat{\sigma}^+_{n_l}\hat{\sigma}^-_{n_l} \right)  \label{eqn:spin_Lindblad}
\end{eqnarray}
Here, $\chi_{n_l,m_k} = 4g_{n_l}g_{m_k}\Delta_c/(4\Delta_c^2 + \kappa^2)$ characterises the strength of the elastic interactions and 
$\Gamma_{n_l,m_k} = 4g_{n_l}g_{m_k}\kappa/(4\Delta_c^2 + \kappa^2)$ the inelastic interactions. In the limit of $\Delta_c \gg \kappa/2$, the Hamiltonian contribution dominates the dynamics and describes an effective XX-Heisenberg model. Physically, the origin of the spin-spin interaction can be understood by the fact that when the optical cavity is detuned from the atomic resonance it enhances the probability that a photon emitted into the cavity by one atom is reabsorbed by another atom before the photon has a chance to leak out through the cavity mirrors. The emission and absorption of the photon realizes the spin exchange process. 

In the main text and the following SOM sections we will generally assume homogeneous interactions $\chi_{n_l,m_k} \equiv \chi$ (and similarly, homogeneous dissipation $\Gamma_{n_l,m_k} \equiv \Gamma$) as this encapsulates the essential physics. Specific cases for which we are interested in the quantitative effects of inhomogeneous interactions are dealt with as required, for example, Sec.~2 of the SOM. Although the atom-light interaction in general depends on the internal hyperfine levels, in this work we deal with situations where only the spatial dependence is relevant. Consequently, we herein simplify the index $n_l \rightarrow l$ with $l$ an integer associated with the $l$-th lattice site, and summations are taken to run over $N$ lattice sites of unit occupation. Whilst the experimental system is composed of $N$ atoms and $\sim 10^3$ lattice sites, such that each site can be occupied by up to $\sim 10^2$ atoms, calculations assuming $N$ lattice sites with unit occupation are equivalent for all quantities considered in the following sections.


\section{SOM Section 2: Mean field treatment of OAT signatures and accounting for inhomogeneity of atom-light coupling \label{sec:SOM_OAT}}
%
%
%
%


In the main text we present evidence of OAT dynamics via a measurement of the frequency shift $\Delta\omega_{\ell}$ of the emitted cavity field, after toggling the cavity detuning from the atomic transition between two frequencies. The observed frequency shift of the emitted radiation can be directly linked to the OAT frequency shift associated with the precession of the atomic coherence. In this section we outline a mean-field model that incorporates all relevant physical processes, including both inhomogeneity of the atom-light coupling and the effects of superradiance.




\subsection{OAT frequency shift for homogeneous atom-light coupling}

We first discuss the origin of the observed OAT shift at the simplest qualitative level, by considering the case of homogeneous interactions.
At the mean-field level the equations of motion for the collective spin variables $J_\alpha= \langle \hat{J}_\alpha \rangle$, with $\alpha=x,y,z$ the orientation of the collective spin, can be written as
\begin{eqnarray}
 \frac{d J^+}{dt} &=&{-i}({2\chi +{i}\Gamma}) J^+(t) J_z(t) \label{eqn:SOM_MF_Jplus}\\
  \frac{d J_z}{dt} &=&  { - \Gamma} J^+(t) J^-(t)\label{eqn:SOM_MF_Jz}
\end{eqnarray} 
Here we assumed  $J_x= Re[J^+]$ and $J_y= Im[J^+]$. Neglecting dissipation for simplicity, $\Gamma = 0$, the equations are simplified as $J_z(t) = J_z(0)$ is a conserved quantity. Consequently, Eq.~(\ref{eqn:SOM_MF_Jplus}) trivially describes a precession of the atomic coherence $J^+(t) = J^+(0)e^{i\omega_{OAT}t}$ where $\omega_{OAT} = -2\chi J_z(0) $ is the OAT frequency shift, which scales linearly with population inversion, and  manifests in the frequency of the emitted superradiant light $\omega_{\ell}$.

\subsection{Effects of inhomogeneous atom-light coupling}

To theoretically describe the inversion-dependent frequency shift observed in Fig.~2b of the main text, one must account for the fact that the coupling of each atom to the cavity mode is not identical. This inhomogeneous coupling must be accounted for both during the dynamical evolution, and during the state preparation step that establishes an initial effective inversion $J_z^\prime$.

In the experiment, the primary source of inhomogeneity arises because the wavelength of the laser at $\lambda_L=813$~nm that creates the standing wave trapping lattice, is incommensurate with the wavelength of the laser light emitted into a cavity mode at $\lambda_c=698$~nm, both forming a standing wave along the Z direction. The atom-light coupling  varies for each lattice site as $g_l \equiv g_0 \mathrm{cos}(k_0 l)$ with $k_0=\pi \lambda_c/\lambda_L$, where $l$ is an integer that labels the $l$th lattice site and $g_0$ is the atom-light coupling at an anti-node of the cavity mode. The inhomogeneity is imprinted in the emitted cavity field [see Eq.~(\ref{eqn:SOM_cavitymode})] which is slaved to the effective atomic coherence as $\hat{a}(t) \propto \hat{\tilde{J}}^-$, where $\hat{\tilde{J}}^{\alpha} \equiv \frac{1}{2g_0}\sum_{l=1}^N g_{l} \hat{\sigma}^{\alpha}_l$ is a generalized collective operator in which each atom's contribution is weighted by its coupling strength to the cavity mode. We highlight for the reader that these new collective variables should not be treated as true spin operators, as they do not form a closed algebra under the usual spin commutation relations: $[\hat{\tilde{J}}^{\alpha}, \hat{\tilde{J}}^{\beta}] \neq i\epsilon_{\alpha\beta\gamma} \hat{\tilde{J}}^{\gamma}$.


We can gain insight into the expected OAT frequency shift in the case of inhomogeneous atom-light coupling by considering the mean-field equation of motion (with $\Gamma = 0$)
\begin{equation}
 \frac{d\tilde{J}^+}{dt} = -i\tilde{J}^+ \sum_{l=1}^N \chi_{ll} \sigma^z_l(t)  \label{eqn:SOM_dSptilde_dt_mf}
\end{equation}
where $\sigma^z_l(t) \equiv \langle \hat{\sigma}^z_l(t) \rangle$. As a coarse approximation we make the assumption $\sigma^z_l(t) \approx \sigma^z_l(0)$ in the absence of dissipation. Whilst this is always true for homogeneous interactions 
(as $\sigma^z_l(0)$ is a conserved quantity), it is only strictly valid for short times in the case of inhomogeneous interactions. 
Nevertheless, it is insightful to solve Eq.~(\ref{eqn:SOM_dSptilde_dt_mf}) with this approximation to estimate the short-time 
frequency shift, yielding for the atomic coherence:
\begin{equation}
 \tilde{J}^+(t) \approx \tilde{J}^+(0) e^{-it\sum_{l=1}^N \chi_{ll}\sigma^z_l(0)}  \label{eqn:SOM_Sptilde_mf}
\end{equation}

The OAT frequency shift observed experimentally via cavity emission can be derived from Eq.~(\ref{eqn:SOM_Sptilde_mf}) by making the mean-field approximation $\langle \hat{a}^{\dagger}(t+\tau) \hat{a}(t) \rangle \equiv \langle \hat{\tilde{J}}^+(t+\tau) \hat{\tilde{J}}^-(t) \rangle \approx \langle \hat{\tilde{J}}^+(t+\tau) \rangle \langle \hat{\tilde{J}}^-(t) \rangle$. Under this assumption we identify $\omega_{OAT} = -\sum_{l=1}^N \chi_{ll}\sigma^z_l(0)$. 

We can define an effective inversion  $J^{\prime}_z$ given by a weighted average over each atom's initial inversion

\begin{equation}
    J^{\prime}_z \equiv \frac{\sum_{l=1}^N g^2_l\sigma^z_{l}(0)/2}{\sum_{l=1}^N g^2_l} \, \label{eq:SOM_Jzprime}
\end{equation}

\noindent as well as an effective interaction strength $\chi^{\prime}$

\begin{equation}
   \chi^{\prime} \equiv \frac{4 \Delta_c}{4\Delta_c^2 + \kappa^2} \left(\frac{1}{N}\sum_{l=1}^N g^2_l\right)
\end{equation}

\noindent such that $\omega_{OAT}=-2 \chi^{\prime}J^{\prime}_z$.  With this definition, $J_z^{\prime}$ can take values between $\pm N/2$. For the specific inhomogeneity present due to the incommensurate cavity mode and lattice wavelengths, $g_l \equiv g_0 \mathrm{cos}(k_0 l)$,  one finds
\begin{equation}
 \chi^\prime \approx \frac{1}{2}\frac{4g_0^2 \Delta_c}{4\Delta_c^2 + \kappa^2} 
 \label{eq:chiprime}
   \end{equation}


\subsection{Experimental observation of OAT frequency shift versus effective inversion}

To determine the inversion-dependent frequency shift of Fig.~2b in the main text, one must calculate $J_z^\prime$ by understanding how the inversion of each atom is initially established.  In the experiment, the atoms are initially prepared in the ground state and then light resonant with the optical transition is injected into the cavity for an amount of time $T$.  An atom initially in the ground state, located at an anti-node of the standing wave cavity mode with characteristic coupling $g_0$ has its polar angle changed by $\theta= \Omega T$, where $\Omega$ is the Rabi frequency associated with injected field for an atom at an antinode of the cavity. An atom with coupling $g_l = g_0 \cos( k_0 l)$ has its polar angle changed by $\theta \cos(k_0 l)$, such that the atom's inversion established by the state preparation pulse is $\sigma^z_l = -\mathrm{cos}[\theta\mathrm{cos}(k_0l)]$.  Using this inversion in Eq.~(\ref{eq:SOM_Jzprime}), one expects that the state preparation pulse creates an effective inversion that depends on the rotation angle as

\begin{equation}
    J_{z,\mathrm{inh}}^{\prime}  = - N\left[\frac{J_1(\theta)}{\theta} - J_2(\theta)\right]  \, 
\label{eqn:jzprime}
\end{equation}

\noindent  Here, $J_{1,2}$ are the first and second order Bessel functions. We have denoted the effective inversion as $J_{z,\mathrm{inh}}^{\prime}$ to differentiate from the effective inversion $J_z^\prime$ extracted from experimental data and plotted in Fig.~\ref{fig:shifts}b of the main text.  Due to the inhomogeneity, full inversion can never be established via the method used here because atoms rotate at different rates in response to the applied state preparation field.

Experimentally, the effective inversion $J_z^{\prime}$ versus $T$ is obtained from a measurement of the total population inversion $J_z$ versus  $T$. This is necessary primarily because $\Omega$ is not known with sufficient accuracy to accurately predict $\theta$, but also to confirm the above model's prediction for the effective inversion established by the state preparation pulse.  We determine $J_z$ by applying 461~nm light to obtain a fluorescence signal proportional to the number of atoms remaining in the ground state just after the state preparation pulse.  The contribution of an atom to this signal is independent of the atom's coupling strength to the cavity mode. The calibration factor relating the fluorescence signal to the total atom number $N$ is established via separate measurements of a collective vacuum Rabi splitting from the $7.5$~kHz linewidth transition versus the fluorescence signal.

For the standing wave cavity mode used to rotate the atoms, we expect $J_z =  -\frac{1}{2}N J_0(\Omega T)$, where $J_0$ is the zeroth order Bessel function.
We observe that the measured $J_z$ versus $T$ shown in Fig.~\ref{fig:supfig}a is well described by $J_z = -\frac{1}{2}N [(1-f) J_0(\Omega T) + f ]$ with fitted parameters $\Omega/2 \pi =1.60(2) $~kHz and $f=0.22(1)$.  The parameter $f$ indicates that some fraction of the atoms are not rotated by the state preparation pulse, perhaps from imperfect spin polarization or finite linewidth of the 698~nm laser used to drive the rotations. 


In a separate set of experiments, we apply the state preparation pulse, and estimate the frequency of the light emitted from the cavity $\omega_\ell$ from the power spectrum of the first 8~ms of light emitted directly after the state preparation pulse.  For a given $T$, we toggle the detuning of the cavity between trials by $\Delta_c/2\pi=\pm 29$~kHz and compute the frequency shift $\Delta\omega_\ell=\omega_\ell|_{\frac{\Delta_c}{2\pi}=+29~\mathrm{kHz}}-\omega_\ell|_{\frac{\Delta_c}{2\pi}=-29~\mathrm{kHz}}$.
In the following we will assume that the effective inversion during the 8~ms of emitted light does not change significantly, supported emperically by varying the window length to shorter and longer periods as well as detailed theoretical modeling in the following section. 

The measured frequency shift $\Delta\omega_\ell$ versus $T$ is shown in Fig.~\ref{fig:supfig}b, and is well described by a fit of the form  $\Delta\omega_{\ell,\mathrm{fit}} =  2 \Delta \chi^{\prime}_{fit} N (1-f)[\frac{J_1(\Omega T)}{\Omega T}-J_2(\Omega T)]) + \omega_{\circ}$, where the coefficients $\Delta\chi^{\prime}_{fit}=\chi^{\prime}|_{\frac{\Delta_c}{2\pi}=+29~\mathrm{kHz}}-\chi^{\prime}|_{\frac{\Delta_c}{2\pi}=-29~\mathrm{kHz}}$ and $\omega_\circ$ are fitted parameters, $f$ and $\Omega$ are from the fit to $J_z$ versus $T$  previously described, and $N$ is the independently measured total atom number.  We find that the data is better described by the inclusion of a fitted constant offset frequency $\omega_\circ$ whose origin is unclear. The fitted value $\omega_\circ/2 \pi = 4(1)  $~Hz is consistent within errors with the frequency offset being computed from $\omega_\circ/2\pi=  2 \Delta\chi^{\prime}_{fit} N f /2 = 4(1)$~Hz. One possible explanation for this apparent coincidence is that the the fraction of atoms $f$ that are not rotated by the state preparation pulse none-the-less contribute fully to the one axis twisting frequency shift.  Why this should be the case is unclear, and it is also possible that the value of the fitted offset frequency is simply an artifact of some variation in a parameter that shifts the lasing frequency in a way that correlates with changes in the cavity resonance frequency, but does so in a way that does not scale with inversion, despite our best attempts to eliminate and identify such a mechanism.

Importantly, the inversion-dependent frequency shift predicted for one-axis twisting does not depend strongly on including this offset frequency or not. The fitted value of the scale factor is  $N \Delta\chi^{\prime}_{fit}/2\pi= 17(3)$~Hz  when $\omega_\circ$  is allowed to be independently fit compared to $N \Delta\chi^{\prime}_{fit}/2\pi= 16(4)$~Hz, when the offset frequency is fixed to zero during the fit.  This is to be compared to a theoretically predicted value $N \Delta\chi_{pred}^{\prime}/2\pi= N(\chi^\prime|_{\frac{\Delta_c}{2 \pi}= 29~\mathrm{kHz}}-\chi^\prime|_{\frac{\Delta_c}{2 \pi}= -29~\mathrm{kHz}})R/2\pi=24(7)~$Hz, where $\chi^\prime$ is computed from Eq.~\ref{eq:chiprime}, and $R=0.92(3)$ is an additional scale factor that accounts for radial spatial averaging of the coupling of the atoms to the cavity mode not reflected in Eq.~\ref{eq:chiprime}.  The error on the prediction is dominated by the uncertainty on the linewidth of the optical transition 1.1(3)~mHz \cite{Ye2007} of approximately $30\%$ leading to the same fractional uncertainty on the predicted value of $g^2_0$.

For visualizing the inversion dependent frequency shift of one-axis twisting, Fig.~\ref{fig:shifts}b in the main text and Fig.~\ref{fig:supfig}c in the supplemental material show parametric plots of the measured $\Delta\omega_\ell$ versus an effective inversion computed from measured and fitted parameters as $J_z^\prime = -N(f/2+ (1-f)[\frac{J_1(\Omega T)}{\Omega T}-J_2(\Omega T)])$ at the same time $T$.  Choosing to compute $J_z^\prime$ in this manner is equivalent to assuming that atoms that do not appear to rotate, nonetheless fully contribute to the one axis twisting dynamics.  However, this choice has no impact on the extracted value of $N\Delta\chi^{\prime}_{fit}$ discussed previously.

\subsection{Impact of superradiance}
The analysis of the frequency shift versus inversion data in the main text and in the previous section assumed that the effective inversion $ J_z^\prime$ did not change during the short 8 ms window during which the experiment measures the frequency of the light. However, at the finite detunings where the experiment operates, superradiant emission can be non-negligible.  As a result, the inversion will change in time and thus $\omega_{OAT}$  as well,  even during the short 8 ms window. Here, we extend our mean-field model to account for superradiance (i.e., $\Gamma \neq 0$). We demonstrate that while inhomogeneous atom-light couplings broadly explain the inversion-dependent linear trend of the experimental data within uncertainties, the interplay of inhomogeneous interactions and dissipation can give insight into understanding the scatter of the experimental results.



\begin{figure*}
 \centering
    \includegraphics[width=18cm]{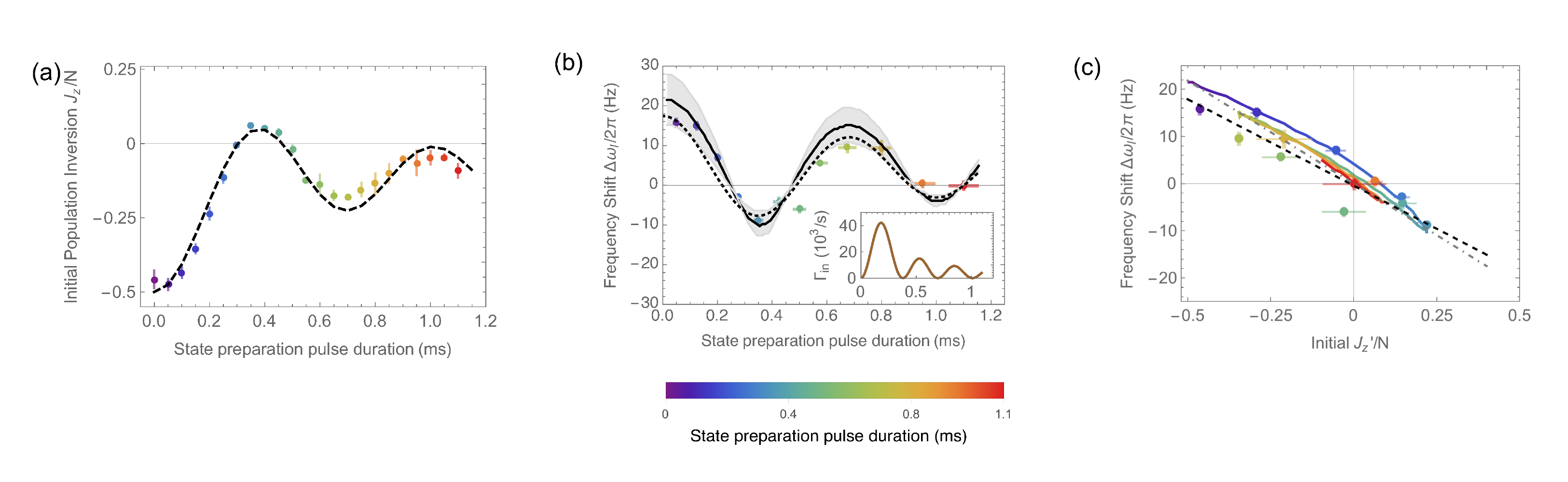} 
  \caption{
  (a) Initial atomic inversion $J_z$ versus duration of the preparation pulse. Experimental data (colored markers) is fitted to an ideal theoretical prediction (dashed line, see main text).
  (b) Frequency shift versus duration of the preparation pulse. Experimental data (colored markers) is compared to the fit  $\Delta\omega_{\ell,\mathrm{fit}} =-  2 \Delta \chi^{\prime}_{fit} J_z^{\prime}$ obtained  from the procedure described in the text (dashed line), and also  to the  numerical simulation of Eqs.~(\ref{eq:sm})-(\ref{eq:sz}), which accounts for both imperfect state preparation and superradiance (solid line). The approximately $\pm30\%$ uncertainty in $\chi^{\prime}$ in this calculation, due to the uncertainty in the known value of the excited state linewidth, is indicated by the gray shaded region.
   Inset: Initial rate of decay of the inversion, $\Gamma_{\mathrm{in}} \propto \langle \hat{\tilde{J}}^+(0) \hat{\tilde{J}}^-(0) \rangle$, as a function of the pulse duration.
  (c) Frequency shift versus the effective inversion $J_z'/N$. The experimental data (colored markers) is compared to the numerical simulation [colored solid line, computed identically to the solid black curve in panel (b)]. A linear fit to the experimental data (black dotted line, also shown in Fig.~2b of the main text) is also plotted, and compared to the ideal  theoretical prediction $\Delta\omega_{\ell} = -2 \Delta\chi_{pred}^{\prime}J_z'$ (gray dashed line, $\Gamma=0$). In all panels the experimental data are colored according to the pulse duration, indicated by the legend. Parameters of the numerical simulation are summarized in the text.}
\label{fig:supfig}
\end{figure*}

Our analysis is based upon a mean field treatment of the master equation [Eq. (2)], which leads to the following equations of motion:
\begin{widetext}
\begin{eqnarray}
\frac{d\sigma_l^+}{dt}&=&=i\frac{\chi}{g_0^2}\sum_{l'=1}^N  (1- \delta_{ll'})g_l g_l\sigma_l^z\sigma_{l'}^++\frac{\Gamma}{2g_0^2}\sum_{l'=1}^N (1- \delta_{ll'}) g_lg_{l'}\sigma_l^z\sigma_{l'}^+ + \frac{i}{g_0^2}\left(\chi + \frac{i\Gamma}{2}\right)g_l^2\sigma_l^+-\gamma_{el}\sigma_l^+ \label{eq:sm}\\
\frac{d\sigma_l^z}{dt}&=&-2i\frac{\chi}{g_0^2}\sum_{l'=1}^N g_lg_{l'}(\sigma_l^+\sigma_{l'}^--\sigma_l^-\sigma_{l'}^+)-\frac{\Gamma}{g_0^2}\sum_{l'=1}^N (1 - \delta_{ll'}) g_lg_{l'}(\sigma_l^+\sigma_{l'}^-+\sigma_l^-\sigma_{l'}^+)-\frac{\Gamma}{g_0^2} g_l^2(\sigma_l^z+1) \label{eq:sz}
\end{eqnarray}
\end{widetext}
Here, we have used that the inhomogeneity of the dissipative contribution is identical to the elastic interactions, $\Gamma_{ll'} \equiv \Gamma g_l g_{l'}/g_0^2$.

Using the same mean-field approximation to associate the observed frequency shift of the cavity emission with the atomic coherence, we numerically solve these equations for many atoms for evolution times corresponding to the $8$~ms measurement time in the experiment. The frequency is then obtained from the Fourier transform of $\tilde J^+(t)$. 

The results of the theoretical simulation of Eqs.~(\ref{eq:sm}) and (\ref{eq:sz}) are directly compared with experimental data in Fig. \ref{fig:supfig}. To ensure our theoretical analysis is consistent with the treatment of the experimental data [Fig.~\ref{fig:shifts}b of the main text and Fig.~\ref{fig:supfig}], we follow an identical procedure to obtain the effective inversion $J_z^{\prime}$ to which we compare the frequency shift. The simulations include both the coupling constant inhomogeneity and the imperfect state preparation. For the latter about $22\%$ of the atoms were randomly selected and kept in the ground state. Parameters used in the numerical simulation were: $N=3.65\times10^5$, $\gamma_{el}=10^2$s$^{-1}$, $\Delta_c=\pm 2\pi\times 30$~kHz, $\kappa = 2\pi \times 145$~kHz and we include a rescaling of the interaction $\chi\rightarrow R\chi$ with $R=0.92$ to characterize radial spatial averaging of the atom-light coupling.

Panel (a) shows the population inversion and (b) the frequency shift for different pulse duration times $T$. We fit the experimental data in each panel according to the method outlined in the previous subsection (indicated by the dashed lines in both panels). The experimentally observed frequency shift in panel (b) is captured well by the numerical calculations including collective dissipation. The uncertainty in the known excited state linewidth $\gamma$ and thus $\chi$ (approximately $30 \% $) can be  incorporated in  the numerical calculations and is indicated by the shaded region.

Panel(c) shows the relationship between the frequency shift and $J_z'/N$. For rotation angles $\theta$ smaller than $\pi$, we observe that the net effect of collective emission is to decrease the population inversion and thus the effective value of $J_z'/N$. In turn, this leads to an associated modification of the observed frequency shift. This effect is dominant close to the equator when  $J_z'/N\sim 0$. For larger $\theta$ the inhomogeneity of the single-particle rotations result in reduced coherence which suppresses the decay of the population inversion. We illustrate this by plotting the initial decay rate as a function of pulse duration in the inset of panel (b). The suppression of superradiance for longer pulses  leads to an observed frequency shift closer to the ideal ($\Gamma = 0$) prediction of $\Delta\omega_{\ell} = -2 \Delta\chi_{pred}^{\prime}J_z'$. The numerical simulations accounting for both inhomogeneity and dissipation reproduce the experimental observations well, confirming the validity of our model.


The effect of superradiant emission  can be roughly estimated by a closer analysis of the mean-field equations of motion. The second term in Eq.~(\ref{eq:sz}) represents the dominant change of inversion coming from collective dissipation. We can crudely approximate the $\sigma_l^z$ used to evaluate the effective inversion $J_z'$  [Eq.~(\ref{eq:SOM_Jzprime})], by an average value accounting for the change due to collective dissipation over the relevant timescale for the frequency shift measurement $\tau=8$~ms: 
$\sum_{l=1}^N \sigma^z_l\approx J_z(0)-\frac{\Gamma\tau}{2Ng_0^2}\sum_{l=1,l'=1}^N g_lg_{l'} [\sigma_{l'}^+(0)\sigma_l^-(0) + c.c.]$, 
where  $\sigma_l^+(0)=\sin(\cos(k_0 l)\theta)/2$, is the initial atomic coherence of those atoms  coupled to the cavity pulse (here we ignore imperfect state preparation for our simple estimate). 
Replacing the summation by a spatial average using a continuous integral of the form $\frac{N}{2\pi}\int_{-\pi}^{\pi} d\theta$, this leads to 
\begin{eqnarray}
 \Delta\omega_{\ell}'(\tau)\approx -4\chi'\left (J_z'-\Gamma N^2\tau\frac{J_1(\theta)^2}{8} \right) \label{eq:effJzdiss}
\end{eqnarray}
The above expression encapsulates the qualitative modifications induced by collective decay to  the frequency shift, although its quantitative validity is limited by our crude treatment of the time evolution to lowest order (linear in $\tau$). An inspection of Fig.~\ref{fig:supfig}c demonstrates that this expression excellently captures the qualitative trend of the experimental data and its deviation from the naive linear prediction $\Delta\omega_{\ell} = -2 \Delta\chi_{pred}^{\prime}J_z'$. In particular, for moderate rotation angles we see a positive increase in the frequency shift due to an effective reduction in the average inversion due to superradiance, consistent with the sign of the contribution $\propto \Gamma$ in the above Eq.~(\ref{eq:effJzdiss}). The decay of the the Bessel function, $J_1(\theta)$, with  increasing pulse area $\theta \gg \pi$,  (due to a reduction in effective coherence) also explains  the more linear scaling of the  OAT shift in this regime.

\section{SOM Section 3: Role of atomic coherence in many-body gap protection and OAT dynamics}


A key point emphasized in the main text is the intrinsic role of the atomic coherence in the dynamics that we observe. More specifically, by rewriting the collective Hamiltonian (implicitly assuming homogeneous interactions herein) as: 
\begin{equation}
\hat{H}_{\mathrm{eff}} = \hbar \chi\hat{J}^+\hat{J}^- \equiv \hbar \chi[\hat{J}^2 - \hat{J}_z^2 + \hat{J}_z]
\end{equation}
we note that the key difference to the generic OAT Hamiltonian is the inclusion of the term $\propto \hat{J}^2$. 

Our experimental observations focus towards demonstrating two outcomes of this additional term: (i) many-body gap protection against slow single-particle noise sources, and (ii) the importance of atomic coherence in generating and observing OAT dynamics. In the following section we outline a theoretical understanding of the experimental observations with respect to these phenomena. 

\subsection{Gap protection}
In Figure 3 of the main text we demonstrated the role of the gap protection for the case of two large collective spins by showing the spectra of the emitted light for different Zeeman energy splitting between the two collective spins. 
We can gain significant insight into this experimental observation by studying the simplified mean-field dynamics of the two collective spins. 

As per the main text, we begin by defining the collective sum, $\hat{J}_z = \hat{J}^1_z + \hat{J}^2_z$, $\hat{J}^{\pm} = \hat{J}^{\pm}_1 + \hat{J}^{\pm}_2$, and difference operators $\hat{j}_z = \hat{J}^1_z - \hat{J}^2_z$, 
$\hat{j}^{\pm} = \hat{J}^{\pm}_1 - \hat{J}^{\pm}_2$ for convenience. The scheme of Fig.~\ref{fig:jdotj} is characterized by the Hamiltonian $\hat{H} = \hbar\chi\hat{J}^+\hat{J}^- + \hbar\delta\hat{j}_z$ where $\hbar\delta$ describes the Zeeman energy splitting between the two collective spins. If we neglect dissipation ($\Gamma = 0$) for simplicity, the associated mean-field equations of motion can be reduced to
\begin{eqnarray}
  \frac{d J^+}{dt} &=& {-i} {2\chi} J^+(t) J_z(0) {+i}\delta j^+(t)  \label{eqn:SOM_Jplus} \\ 
  \frac{d j^+}{dt} &=& {-i} {2\chi } J^+(t) j_z(t) {+i}\delta J^+(t)  \label{eqn:SOM_jplus} \\ 
  \frac{d^2 j_z}{dt^2} &=&  - 4\left[\delta^2 + J_z^2(0) + J_{\perp}^2(0)\right]j_z \notag \\
    & & + 6\delta\chi j^2_z + 2\delta\chi J_{\perp}^2(0)  \label{eqn:SOM_jz}
\end{eqnarray}
where the total inversion $J_z(t) \equiv J_z(0) = -N\mathrm{cos}(\theta)/2$ is a conserved quantity, $J_{\perp}^2(0)\equiv |J^+(0)|^2$, $\theta$ is the rotation of the initial state from the south pole of the Bloch sphere and we consider 
a total of $N/2$ atoms in each ensemble. 


To solve for the dynamics, the generic approach is to first solve for $j_z(t)$ and substitute the solution into Eqs.~(\ref{eqn:SOM_Jplus}) and (\ref{eqn:SOM_jplus}). 
For simplicity, we consider an approximate solution for $j_z(t)$ wherein we treat Eq.~(\ref{eqn:SOM_jz}) as an equation of motion describing a classical particle in a potential. 

More specifically, we use that $\ddot{j_z} \equiv -dU(j_z)/dj_z$ describes motion of a classical particle in the potential 
{\small \begin{eqnarray}
 U(j_z) = -2\delta\chi J_{\perp}^2(0) j_z + 2\left[ \delta^2 + J_z^2(0) + J_{\perp}^2(0) \right]j_z^2 -2\delta\chi j_z^3 \notag
\end{eqnarray}}
In this formalism, the prepared state of the experimental system has an initial  (conserved) energy $E(0) \equiv U(j_z(0)) + 1/2 \dot{j_z}(0)^2$ (setting an effective mass equal to one). As $j_z(0) = 0$ and $\dot{j_z}(0) = -2i\chi[J^+_1(0)J^-_2(0) - J^+_2(0)J^-_1(0)] = 0$ for $J^+_1(0)=J^+_2(0)$ initially, we have that $E(0) =0$. Our approximation lies in identifying that the potential, whilst formally cubic, can be well approximated as a harmonic potential 
by using the Taylor expansion
\begin{equation}
 U(j_z) \approx U(j_0) + \frac{1}{2}\left.\frac{d^2U}{dj^2_z}\right\vert_{j_z = j_0} (j_z - j_0)^2  \label{eqn:SOM_HarmonicApprox}
\end{equation}
about the local minima
\begin{eqnarray}
 j_0 = \frac{-\delta^2 - \chi^2 N^2 - \sqrt{(\delta^2 + \chi^2N^2)^2 - 3\delta^2\chi^2n^2\mathrm{sin}(\theta)^2}}{3\delta\chi} \notag
\end{eqnarray}

Using this harmonic approximation, the dynamics of the classical particle can be trivially solved to yield the general solution 
\begin{widetext}
\begin{eqnarray}
 j_z(t) \simeq \mathcal{A}_0\mathrm{sin}^2(\omega_0 t) \label{eqn:SOM_jz_soln} \\[8pt]
 \mathcal{A}_0 = \frac{4\delta^2 + \chi^2N^2 - \sqrt{(4\delta^2 + \chi^2N^2)^2 - 12\delta^2\chi^2N^2\mathrm{sin}^2(\theta) }}{6\delta\chi}  \label{eqn:SOM_A0} \\[8pt]
 \omega_0 = \frac{1}{2}\left[ (4\delta^2 + \chi^2N^2)^2 - 12\delta^2\chi^2N^2\mathrm{sin}^2(\theta) \right]^{1/4}  \label{eqn:SOM_omega0}
\end{eqnarray}
\end{widetext}
where we have used the initial conditions $J_z(0) = -N\mathrm{cos}(\theta)$ and $|J^+(0)| = N\mathrm{sin}(\theta)$. 


In the main text we assume that near the poles of the Bloch sphere we can neglect the contribution of $j_z(t)$ in Eq.~(\ref{eqn:SOM_jplus}), specifically that $\chi j_z(t) \ll |\delta|$. We can justify this approximation by splitting our argument into two limits: (i) $|\delta| \ll \chi N$, and (ii) $|\delta| \gtrsim \chi N$. In the former case we have that oscillations in $j_z(t)$ are bounded by $|j_z(t)| \leq |\delta|/\chi\mathrm{sin}^2(\theta)$, whilst in the latter we instead have the bound $|j_z(t)| \leq (N/2)\mathrm{sin}^2(\theta)$. In both cases the limit $\chi j_z(t) \ll |\delta|$ follows from application of $\mathrm{sin}^2(\theta) \ll 1$ and the relevant inequalities defining the relative scale of $\delta$ and $\chi N$.

Using this justification we can then map the problem into a linear two level system with effective detuning $\chi N$ and off diagonal coupling $\delta$. When $\chi N >\delta$, the large effective detuning prevents population transfer between the levels.

\begin{figure}
 \centering
    \includegraphics[width=3.375in]{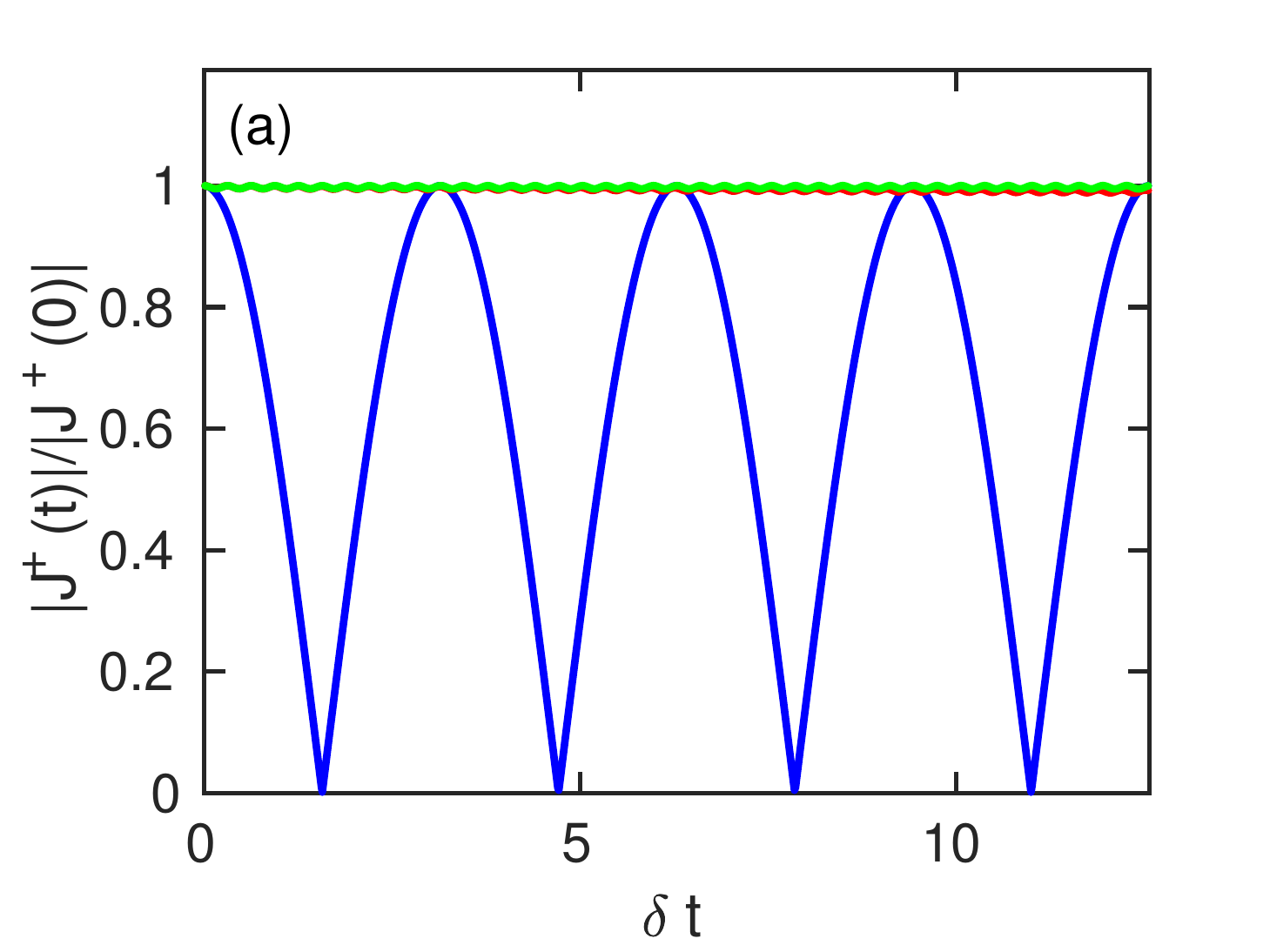} 
    \includegraphics[width=3.375in]{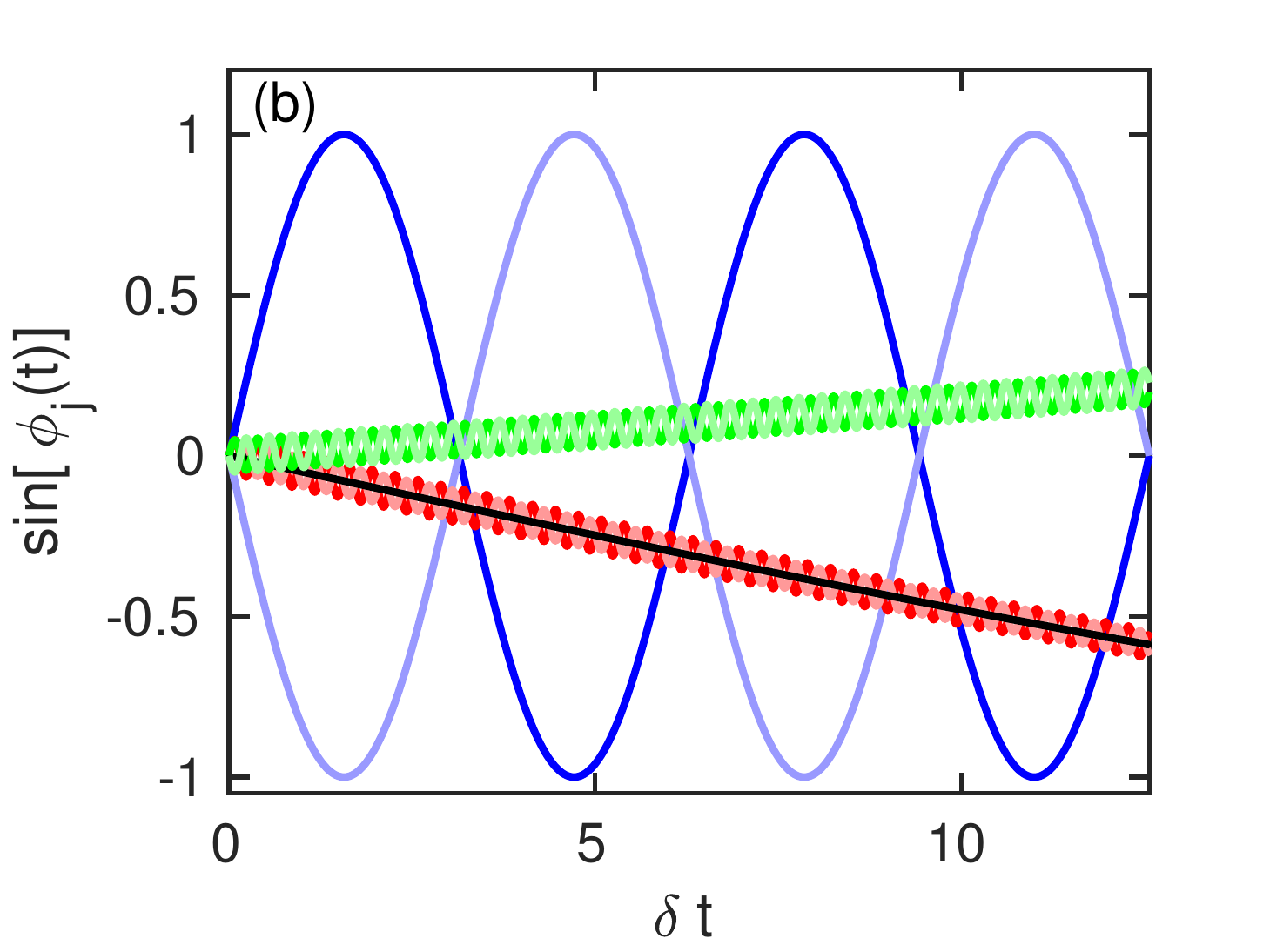}
  \caption{Amplitude of the total atomic coherence $J^+(t) \equiv |J^+(t)|e^{i\phi(t)}$ [panel (a)] and phase  of the individual atomic coherence $J^+_j(t) \equiv |J^+_j(t)|e^{i\phi_j(t)}$ for $j=1,2$ [panel (b)] in the ideal ($\delta =0$) OAT frame such that $\phi_j(t) \rightarrow \phi_j(t) - \omega_{OAT}$ for $\omega_{OAT} = - 2\chi J_z(0)$. We compare the results of numerical simulation of Eqs.~(\ref{eqn:SOM_Jplus})-(\ref{eqn:SOM_jz}) for initial rotation angle $\theta = \pi/10$ (red data) and $\theta = 6\pi/10$ (green data) to that obtained for the generic OAT Hamiltonian (blue lines, see text). For the state prepared near the south pole we find the phases of the two spins track the simplified prediction of $\phi_j(t) \sim \omega_+ t$ (black line) where $\omega_+ = \left(\chi N + \sqrt{4\delta^2 + \chi^2N^2} \right)/2$. Faint and strong lines indicate the individual collective spins. 
}\label{fig:SOM_GapProtection}
\end{figure}

Although we focused of the case when the initial state is close to the south pole, which is the case relevant for our experiment, it is important to emphasize that the gap protection (at least for the ideal case when we can ignore dissipation,$\Gamma=0$) is effective for all initial rotation angles. This is illustrated in Fig.~\ref{fig:SOM_GapProtection}, for states prepared both near the south pole ($\theta=\pi/10$) and the equator ($\theta=6\pi/10$). We numerically solve Eqs.~(\ref{eqn:SOM_Jplus})-(\ref{eqn:SOM_jz}) and plot the amplitude of the total atomic coherence $J^+(t) \equiv |J^+(t)|e^{i\phi(t)}$ and phase of the individual coherence,$\phi_j(t)$, defined from  $J^+_j(t) \equiv |J^+_j(t)|e^{i\phi_j(t)}$ for $j=1,2$. For clarity, we compare against the dynamics of the generic OAT Hamiltonian $\hat{H} = \hbar\chi\hat{J}^2_z + \hbar\delta\hat{j}_z$. We find that for $|\delta| \ll \chi N$ the amplitude of the total coherence remains almost fixed, whilst the individual collective spins remain locked together and precess at a rate only slightly perturbed from the ideal ($\delta = 0$) OAT frequency shift $\omega_{OAT} = -2\chi J_z(0)$. We find little qualitative difference whether the state is prepared near or far from the south pole, although the evolution of $j_z(t)$ is important for the dynamics of the latter. For the state prepared near the south pole we find good agreement with our simplified solution of the two level system (Eqs. 2 and 3 in the main text with $j_z$ set to zero).


\subsection{Role of atomic coherence in OAT dynamics and measurement of the many-body gap}
Figure \ref{fig:dephase} of the main text illustrates the results of a protocol that controls the atomic coherence to directly spectroscopically probe the many-body energy gap. In this section we give a theoretical underpinning of these results. We begin by outlining a simple mean-field model to describe the dephasing/rephasing protocol of Fig.~\ref{fig:dephase} which accounts for the experimentally observed reduction in the OAT frequency shift. Secondly, we provide a quantum treatment of the problem, as introduced schematically in Fig.~\ref{fig:dephase}~(c), which allows us to elucidate the connection to the many-body gap opened by the $\hat{J}^2$ term of the effective Hamiltonian. 

\subsubsection{Mean-field model}
The simplified mean-field model of two collective spins also gives excellent insight into the role of atomic coherence in the OAT dynamics. Specifically, the dephasing/rephasing of the atomic ensemble can be understood as a coupling between bright and dark modes (referred to as $B_+$ and $B_-$ in the main text) of the two-spin model, and  dynamical evolution of these  modes during the holding period. 


We begin by first considering the dephasing of the atomic ensemble, where the bright and dark modes are initially associated with $J^+(0)$ and $j^+(0)$ observables respectively. In the context of the 
collective spins the dephasing corresponds to $\delta/\chi N \gg 1$ in Eqs.~(\ref{eqn:SOM_Jplus}) 
and (\ref{eqn:SOM_jplus}), such that one realizes an effective beam-splitter for the $J^+$, $j^+$ observables: $J^+(\tau) = \mathrm{cos}(\delta\tau)J^+(0)$ and $j^+(\tau) = i\mathrm{sin}(\delta\tau)J^+(0)$ where $\tau$ is the duration of the dephasing.
Both $J_z=J_z(0)$ and $j_z=0$ are conserved quantities in this limit.

Subsequent to the dephasing of the ensemble, the atoms are held for a duration $T_{hold}$, corresponding to evolution according to Eqs.~(\ref{eqn:SOM_Jplus})-(\ref{eqn:SOM_jz}) with $\delta = 0$. This gives
\begin{equation}
 j_z(t) = \frac{|J^+(0)|^2\mathrm{sin}(2\delta\tau)}{2\sqrt{|J_z|^2 + |J^+(\tau)|^2}} \mathrm{sin}(2\chi\sqrt{|J_z|^2 + |J^+(\tau)|^2}t) 
\end{equation}
Similarly, we can solve
\begin{equation}
 J^+(t) = J^+(\tau)e^{-2i\chi J_z t} 
\end{equation}
Clearly from this last expression, we can identify that the OAT dynamics leads to a frequency shift in the bright mode $J^+$ which is independent of the magnitude of the atomic coherence $|J^+|$. 

To solve for the rephasing dynamics we make the assumption that the system is prepared away from the equator and is sufficiently dephased ($\delta \tau \simeq \pi/2$) such that $|J_z| \gg |J^+(\tau)|$. 
After the final rephasing period ($\delta/\chi N \gg 1$) we then find
\begin{equation}
 J^+(t) = J^+(0)\left[ \mathrm{cos}^2(\delta\tau) e^{-2i\chi J_z t} + \mathrm{sin}^2(\delta\tau) e^{i\frac{|J^+(\tau)|^2}{J_z}t} \right]  \label{eqn:RephasedJp}
\end{equation}
The structure of this solution has a clear implication for OAT dynamics. For large dephasing, $\delta\tau \simeq \pi/2$, the majority of the atomic ensemble occupies the dark mode $j^+$ during the hold 
time and oscillates at a much reduced rate (relative to the expected OAT oscillation frequency shift). Upon rephasing it is this contribution which is dominant in Eq.~(\ref{eqn:RephasedJp}) and characterizes the observed cavity emission.

\subsubsection{Quantum model}
A quantum description of the dephasing/rephasing dynamical protocol discussed in the main text (see Fig.~\ref{fig:dephase} and surrounding text) can also be derived for the case of a collective spin state prepared very close to the south pole. This description rigorously justifies the intuitive interpretation of the dynamics laid out in the schematic picture of Fig.~\ref{fig:dephase}~(c) of the main text. The key assumption of the graphical picture, which we justify herein, is that dephasing ideally maps the initial maximally symmetric state to one of minimal coherence. This assumption subsequently allows us to interpret the observed vanishing OAT frequency shift as a spectroscopic measurement of the many-body energy gap induced by the $\hat{J}^2$ contribution of the Hamiltonian.

We begin our analysis by noting that, due to the conservation of total inversion throughout the dynamics, we can simplify the calculation by considering only the three highest Dicke manifolds $J=N/2$, $N/2-1$ and $N/2-2$. 
Under this approximation, we write the initial spin state in the usual Dicke basis $\vert J,m_J\rangle$ as 
\begin{eqnarray}
 |\psi(0)\rangle & = & \alpha\vert N/2,-N/2\rangle +\beta\vert N/2,-N/2+1\rangle  \notag \\
 & & + \gamma\vert N/2,-N/2+2\rangle
\end{eqnarray}
where the coefficients $\alpha,\beta,\gamma$ (satisfying normalization requirements) are  determined from the formal expansion of the initial coherent spin-state in the Dicke basis. We can calculate the subsequent evolution by treating each magnetization component $m_J$ seperately.

Firstly, the $\vert N/2,-N/2\rangle$ state evolves as, 
\begin{multline}
 e^{i\hat{H}_d\tau/\hbar} e^{-i\hat{H}t/\hbar} e^{-i\hat{H}_d\tau/\hbar}\vert N/2,-N/2\rangle \\
  = e^{-i(E^J_{N/2} - E^z_{-N/2})t/\hbar} \vert N/2,-N/2\rangle  \label{eqn:SOM_Dicke0}
\end{multline}
where $\hat{H}_d = \hbar\sum_n h_n (1+\hat{\sigma}^z_n)/2$ characterizes the single-particle dephasing and $\hat{H} = \hbar\chi(\hat{J}^2 - \hat{J}^2_z + \hat{J}_z)$ the interaction Hamiltonian. We also define $E^J_j = \hbar\chi j(j + 1)$ and $E^z_{m_J} = \hbar\chi (m_J^2-m_J)$ as the energy associated with the $\hbar\chi\hat{J}^2$ and $\hbar\chi\hat{J}^2_z$ terms of the interaction Hamiltonian respectively. As illustrated intuitively in Fig.~\ref{fig:dephase}~(c) of the main text, dephasing cannot couple the initial $\vert N/2,-N/2\rangle$ state to any other Dicke manifold due to conservation of inversion, leading to this simplistic evolution identical to that expected under solely the XX Hamiltonian.

In contrast, the initial states with one ($\vert N/2,-N/2+1\rangle$) or two ($\vert N/2,-N/2+2\rangle$) excitations can couple to the $N/2-1$ and $N/2-2$ manifolds respectively via the dephasing. To evaluate their evolution explicitly we must go beyond the abbreviated Dicke basis $|J,m_J\rangle$ and introduce the spin-wave states:
\begin{eqnarray}
 \vert k \rangle_1 &=& \frac{1}{\sqrt{N}} \sum_J e^{ikj} \hat{\sigma}^+_j \vert N/2,-N/2\rangle  \\
 \vert k k' \rangle_2 & \approx & \frac{1}{N} \sum_{jj'} e^{ikj} e^{ik'j'} \hat{\sigma}^+_j \hat{\sigma}^+_{j'} \vert N/2,-N/2\rangle 
\end{eqnarray}
where $k = 2\pi n / N$ for $n = 1,2,...N-1$. For convenience we adopted the notation  $|0\rangle_1 \equiv |N/2,-N/2+1\rangle$ and $|00\rangle_2 \equiv |N/2,-N/2+2\rangle$ for the fully symmetric Dicke states and refer to  these as `$k=0$' contributions in the relevant summations, but we note they are not spin waves states. We highlight that the definition of the state $|k k'\rangle_2$ is only approximately valid as we spuriously include the contribution $j = j'$ to the summation for simplicity, although it does not affect the insight of our solution. The spin-wave states allow us to fully characterize the evolution of the system under dephasing, which transfers population out of the fully collective subspace. 

A useful result, which we use in the following, is the evolution of the above states under the XX Hamiltonian:
\begin{widetext}
\begin{eqnarray}
 e^{-i\hat{H}t/\hbar} \vert k \rangle_1 & = & e^{i E^z_{-N/2+1}t}\left[\delta_{k,0} e^{-iE^J_{N/2} t/\hbar} +  (1-\delta_{k,0})e^{-i(E^J_{N/2-1} t/\hbar} \right \vert k \rangle_1 , \label{eqn:SOM_kevo_XX}\\
 e^{-i\hat{H}t/\hbar} \vert k k' \rangle_2 & = &  e^{i E^z_{-N/2+2}t}\left[ \delta_{k,0}\delta_{k',0}e^{-iE^J_{N/2} t/\hbar} + (\delta_{k,0}+\delta_{k',0})e^{-iE^J_{N/2-1} t/\hbar} + \right. \notag \\
 & & \left. + (1 - \delta_{k,0}\delta_{k',0})(e^{-iE^J_{N/2-2} t/\hbar} - e^{-iE^J_{N/2-1} t/\hbar})\right] \vert k k' \rangle_2 \label{eqn:SOM_kkevo_XX}
\end{eqnarray}
\end{widetext}
where $\delta_{ij}$ is the Kronecker Delta function. Using the spin-wave basis and Eqs.~(\ref{eqn:SOM_kevo_XX}) and (\ref{eqn:SOM_kkevo_XX}) we are then ready to calculate the evolution of the $\vert N/2,-N/2+1\rangle$ and $\vert N/2,-N/2+2\rangle$ states through each step of the dephasing/rephasing protocol. 

For the initial dephasing step these states evolve as:
\begin{eqnarray}
 e^{-i\hat{H}_d\tau}\vert N/2,-N/2+1\rangle & = & \frac{1}{\sqrt{N}}\sum_n e^{-ih_nt} \vert n \rangle  \label{eqn:SOM_1evo_dephase} 
\end{eqnarray}
and
\begin{eqnarray}
 e^{-i\hat{H}_d\tau}\vert N/2,-N/2+2\rangle & \approx & \frac{1}{N}\sum_{nm} e^{-ih_nt}e^{-ih_mt} \vert n,m \rangle \label{eqn:SOM_2evo_dephase}  
\end{eqnarray}
Here, we introduce the short-hand notation $|n\rangle \equiv \hat{\sigma}^{+}_n|N/2,-N/2\rangle$ and $|n,m\rangle \equiv \hat{\sigma}^{+}_n \hat{\sigma}^{+}_m|N/2,-N/2\rangle$ where $\hat{\sigma}^+_{n}$ is the usual spin-raising operator which creates a spin excitation at the $n$th lattice site. Again, we include the spurious $n=m$ contribution for simplicity of calculation. These states serve as the position-space Fourier partner of the spin-waves.

We can verify that the dephasing maps the states from the maximally symmetric manifold $J=N/2$ to that of minimal coherence for a given $m_J$ by expanding Eqs.~(\ref{eqn:SOM_1evo_dephase})-(\ref{eqn:SOM_2evo_dephase}) in terms of the spin-wave states. Defining $\epsilon = \sum_n e^{-i h_n \tau}/N$ to characterize the degree to which the ensemble is dephased, we can show that for perfect dephasing, $\epsilon\rightarrow 0$, we can express the dephased states [Eqs.~(\ref{eqn:SOM_1evo_dephase})-(\ref{eqn:SOM_2evo_dephase})] as: 
\begin{widetext}
\begin{eqnarray}
 e^{-i\hat{H}_d\tau}\vert N/2,-N/2+1\rangle\vert_{\epsilon=0} = \frac{1}{N} \sum_{\substack{n \\ q\neq0}} e^{-ih_nt}e^{-iqn}\vert q \rangle_1
\end{eqnarray}
and
\begin{eqnarray}
 e^{-i\hat{H}_d\tau}\vert N/2,-N/2+2\rangle\vert_{\epsilon=0} \approx \frac{1}{N^2} \sum_{\substack{nm \\ q\neq0,q'\neq0}} e^{-i(h_n+h_m)t}e^{-iqn - iqm}\vert q,q' \rangle_2
\end{eqnarray}
\end{widetext}
As the spin-wave states $\vert q \rangle_1$ and $\vert q,q' \rangle_2$ occupy the $J=N/2-1$ and $J=N/2-2$ manifolds respectively, we thus have that the fully symmetric initial state is ideally mapped to the manifold of minimal coherence for a given $m_J$, with other contributions vanishing. We note that our definition of the spin-wave state $|q,q'\rangle_2$ implies there will be minor corrections for the initial $\vert N/2,-N/2+2\rangle$ state, making this statement only approximately true.

Returning to the full protocol, we evaluate the subsequent evolution of the dephased states [Eqs.~(\ref{eqn:SOM_1evo_dephase})-(\ref{eqn:SOM_2evo_dephase})] under the XX Hamiltonian by again transforming to the spin-wave states and using the results of Eqs.~(\ref{eqn:SOM_kevo_XX}) and (\ref{eqn:SOM_kkevo_XX}):
\begin{widetext}
\begin{eqnarray}
 e^{-i\hat{H}t}e^{-i\hat{H}_d\tau}\vert N/2,-N/2+1\rangle & = & \frac{e^{-i(E^J_{N/2} - E^z_{-N/2+1})t/\hbar}}{N}\sum_{n,q} e^{-ih_nt/}e^{-iqn}  \left[ e^{-i\Delta_1t} - \delta_{0,q}\left(e^{-i\Delta_1t} - 1\right) \right]  \vert q \rangle_1  \notag  
\end{eqnarray}
 and
\begin{eqnarray}
 e^{-i\hat{H}t}e^{-i\hat{H}_d\tau}\vert N/2,-N/2+2\rangle & \approx & \frac{e^{-i(E^J_{N/2} - E^z_{-N/2+2})t/\hbar}}{N^2}\sum_{\substack{nm \\ qq'}} e^{-ih_nt}e^{-ih_mt} e^{-iqn} e^{-iq'm} \Big[ e^{-i\Delta_2t} - \delta_{0,q}\left(e^{-i\Delta_2t} - e^{-i\Delta_1t}\right) \label{eqn:SOM_1evo_XX} \\
 & & - \delta_{0,q'}\left( e^{-i\Delta_2t} - e^{-i\Delta_1t} \right) - \delta_{0,q}\delta_{0,q'}\left( 2e^{-i\Delta_1t} - e^{-i\Delta_2t} - 1 \right) \Big] \vert q q' \rangle_2 \label{eqn:SOM_2evo_XX} 
\end{eqnarray}
\end{widetext}
where we introduce $\Delta_n = (E^J_{N/2 - n} - E^J_{N/2})/\hbar$ for brevity.

At this point in the protocol, after only dephasing and evolution under the XX Hamiltonian, one can use Eqs.~(\ref{eqn:SOM_1evo_XX}) and (\ref{eqn:SOM_2evo_XX}) in combination with the simple evolution of the state $\vert N/2,-N/2\rangle$, to show that we obtain a OAT frequency shift $\langle \hat{J}^+ \rangle = \langle \hat{J}^+(\tau)\rangle e^{i\chi Nt}$ with $\langle \hat{J}^+(\tau) \rangle = (\beta^*\alpha + 2\gamma^*\beta)\sqrt{N}\epsilon^*$ the coherence of the dephased state [i.e., using Eqs.~(\ref{eqn:SOM_1evo_dephase})-(\ref{eqn:SOM_2evo_dephase})] and $\epsilon$ defined as previous.
Consistent with the results of the prior mean-field calculation, this result demonstrates that the coherence still precesses at a rate $\omega_{OAT} \approx -2\chi \langle \hat{J}_z \rangle$ consistent with the usual OAT result.

The final rephasing step is calculated by transforming back to the position basis $|n'\rangle$ and $|n',m'\rangle$. By appropriately collapsing the resulting summations using the identity $\sum_q e^{iq(n-m)} = N\delta_{n,m}$ and similar, we have the final results
for the evolution of the $\vert N/2,-N/2+1\rangle$ and $\vert N/2,-N/2+2\rangle$ contributions:
\begin{widetext}
\begin{eqnarray}
 e^{i\hat{H}_d\tau} e^{-i\hat{H}t} e^{-i\hat{H}_d\tau}\vert N/2,-N/2+1\rangle & = & e^{-i(E^J_{N/2} - E^z_{-N/2+1})t/\hbar} \left[ e^{-i\Delta_1 t} 
  + |\epsilon|^2 \left( 1 - e^{-i\Delta_1 t} \right) \right] \vert N/2,-N/2+1\rangle \notag \\
 & & + \epsilon\sum_{k\neq 0} c_k |k\rangle_1 \label{eqn:SOM_Dicke1}
\end{eqnarray}
and
\begin{eqnarray}
 e^{i\hat{H}_d\tau} e^{-i\hat{H}t} e^{-i\hat{H}_d\tau}\vert N/2,-N/2+2\rangle 
 & \approx & e^{-i(E^J_{N/2} - E^z_{-N/2+2})t/\hbar} \left[ e^{-i\Delta_2 t} + 2|\epsilon|^2 \left( e^{-i\Delta_1 t} - e^{-i\Delta_2 t} \right) \right. \notag \\
 & & \left. + |\epsilon|^4\left( 1 + e^{-i\Delta_2 t} - 2e^{-i\Delta_1 t} \right) \right] \vert N/2,-N/2+2\rangle 
 + \epsilon \sum_{\substack{k\neq0 \\k'\neq0}}  c_{k,k'}  |k k'\rangle_2 \label{eqn:SOM_Dicke2}
\end{eqnarray}
\end{widetext}
where we neglect the form of the coefficients $c_k$ and $c_{k,k'}$ of the remaining spin-wave terms as they do not contribute to the observable $\langle \hat{J}^+ \rangle$.



In the case of complete dephasing ($\epsilon = 0$) of the atomic ensemble the quantum dynamics thus have an intuitive interpretation. The initial $\vert N/2,-N/2+1\rangle$ and 
$\vert N/2,-N/2+2\rangle$ states are coupled completely to (and back from) the minimally coherent $J=N/2-1$ and $J=N/2-2$ manifolds respectively by the dephasing/rephasing. Consequently, during evolution under $\hat{H}$ they accrue a phase-difference proportional to the 
appropriate energy gap $\Delta_{1,2}$ induced by the $\hat{J}^2$ contribution in the Hamiltonian. This is precisely the physics illustrated in Fig.~4c of the main text and described in the associated caption. 

Substituting $\Delta_1 = -\chi N$ and $\Delta_2 = -2\chi (N - 1)$ it is straightforward to show that for $\epsilon = 0$
\begin{eqnarray}
 \langle \hat{J}^+ \rangle \simeq \langle \hat{J}+(0) \rangle  
\end{eqnarray}
where $\langle \hat{J}+(0) \rangle = \left(\beta^*\alpha + \sqrt{2}\gamma^*\beta\right)\sqrt{N}$ is the initial atomic coherence of the prepared state near the south pole.
The slow rotation of $\langle \hat{J}^+ \rangle$ with frequency $-2\chi$ is emphasized by contrasting to the usual OAT result (as outlined in Sec.~2 of the SOM) which predicts a frequency shift of magnitude $2\chi \langle \hat{J}_z \rangle \approx \chi N$ near the south pole. Such a slowing of phase accrual is consistent with the mean-field picture, and is effectively a spectroscopic probe of the many-body energy gap.

\end{document}